\documentclass[aps,amsmath,amssymb,preprint, 12pt]{revtex4-1}
\usepackage{amsmath, latexsym, color, graphicx, tabularx, booktabs, enumerate}
\usepackage{multirow}
\usepackage{dcolumn}% Align table columns on decimal point
\usepackage{bm}% bold math
\usepackage{hyperref}
\usepackage{mathptmx}
\usepackage{epstopdf}
\usepackage{enumerate}
\usepackage{subcaption}

\usepackage{titlesec}
\titlespacing*{\section}{0pt}{1.1\baselineskip}{\baselineskip}
\date{\today}

\usepackage[normalem]{ulem}

\begin{document}

\author{Niraj K. Nepal$^{1}$}
\email{Corresponding author: niraj.nepal@temple.edu}
\author{Santosh Adhikari$^1$}
\author{Bimal Neupane$^1$}
\author{Adrienn Ruzsinszky$^1$}

\affiliation{[1] Department of Physics, Temple University, Philadelphia, Pennsylvania 19122, United States}
%%%%%%%%%%%%%%%%%%%%%%%%%%%%%%%%%%%%%%%%%%%%%%%%%%%%%%%%%%%%%%%%%%%%%
%% The document title should be given as usual. Some journals require
%% a running title from the author: this should be supplied as an
%% optional argument to \title.
%%%%%%%%%%%%%%%%%%%%%%%%%%%%%%%%%%%%%%%%%%%%%%%%%%%%%%%%%%%%%%%%%%%%%
\title{Formation energy puzzle in intermetallic alloys: Random phase approximation fails to predict accurate formation energies}

%%%%%%%%%%%%%%%%%%%%%%%%%%%%%%%%%%%%%%%%%%%%%%%%%%%%%%%%%%%%%%%%%%%%%
%% Some journals require a list of abbreviations or keywords to be
%% supplied. These should be set up here, and will be printed after
%% the title and author information, if needed.
%%%%%%%%%%%%%%%%%%%%%%%%%%%%%%%%%%%%%%%%%%%%%%%%%%%%%%%%%%%%%%%%%%%%%

%%%%%%%%%%%%%%%%%%%%%%%%%%%%%%%%%%%%%%%%%%%%%%%%%%%%%%%%%%%%%%%%%%%%%
%% The manuscript does not need to include \maketitle, which is
%% executed automatically.
%%%%%%%%%%%%%%%%%%%%%%%%%%%%%%%%%%%%%%%%%%%%%%%%%%%%%%%%%%%%%%%%%%%%%


\begin{abstract}
 	\noindent  We performed density functional calculations to estimate the formation energies of intermetallic alloys. We used two semilocal approximations, the generalized gradient approximation (GGA) by Perdew-Burke-Ernzerhof (PBE) and the strongly constrained and appropriately normed (SCAN) meta-GGA. In addition, we utilized two nonlocal DFT functionals, the hybrid HSE06, and the state-of-the-art random phase approximation (RPA). The nonlocal functionals such as HSE06 and RPA yield accurate formation energies of binary alloys with completely-filled d-band metals, where semilocal functionals underperform. The accuracy at the nonlocal functionals is greatly reduced when a partially-filled d-band metal is present in an alloy, while PBE-GGA outperforms in these cases. We show that the accurate prediction of formation energies by any DFT method depends on its ability to predict the accurate electronic properties, e.g., valence d-band contribution to the density of states (DOS). The SCAN meta-GGA often corrects the PBE-DOS, however, it does not provide accurate formation energies compared to PBE. This is assumed to be due to the lack of proper error cancellation that should be expected due to the similar bulk nature of both alloys and their constituents, which may improve with the modification of meta-GGA ingredients. RPA yields too negative formation energies of alloys with partially-filled d-band metals. RPA results can be corrected by restoring the exchange-correlation kernel, thereby improving the short-range electron-electron correlation in metallic densities.
\end{abstract}


\maketitle
\section{Introduction}
Intermetallic alloys, composed of two or more d-band transition metals, are often interesting for applications, therefore, their application-governed aspects have been mostly explored. In most cases, alloys were classified with respect to various factors such as the radius ratio of two constituents, electronegativity, principle quantum number, ionicity, coordination number, etc. \cite{P07,MP59,SPS04,W18}. However, there is a scarcity of sufficient information about the chemical bonding and its relation with the equilibrium properties such as formation energy in both theory and the experiment \cite{P07}. Recently in 2014, Zhang \textit{et.al.} \cite{ZKW14}, using DFT calculations \cite{HK64,KS65} for the copper-gold alloys, showed that the accurate prediction of formation energy accompanies an accurate prediction of the density of states. Here, we will generalize this result using a more diverse set of compounds. \\

Intermetallic alloys always have been critical tests for various approximations within density functional theory (DFT). The accurate prediction of basic equilibrium properties of intermetallic alloys and their bulk transition metal constituents with many popular DFT approximations is difficult. Semilocal approximations such as the local density approximation (LDA) \cite{KS65} and various generalized gradient approximations (GGA) are unable to provide accurate formation energy (heat of formation, E$_f$) of weakly-bonded (WB) systems such as copper-gold alloys \cite{OWZ98,ZKW14,TLRKNV16,NABR19}. Incorporating an amount of nonlocality by the kinetic energy density ($\tau(r)$), meta-GGAs slightly improve the equilibrium properties including formation energies of copper-gold alloys \cite{NABR19}, but fail to improve beyond PBE \cite{PBE96} when dealing with the more strongly-bonded (SB) systems such as HfOs and PtSc \cite{IW18}. Attempts at correcting the semilocal results with zero-point vibration energy, additive long-range van der Waals (vdW) interaction, and spin-orbit coupling couldn't improve the result for copper-gold alloys \cite{NABR19}.\\

In general, hybrid functionals within the generalized Kohn-Sham theory that mix the non-local exact exchange with DFT exchange do not provide accurate equilibrium properties of bulk transition metals \cite{JLSVL14}. However, surprisingly, the hybrid HSE06 \cite{HSE03,HSE03-er} shows some promise in the prediction of the lattice constants and formation energies of WB systems \cite{ZKW14}. It couldn't provide a reasonable bulk modulus, which is a fundamental physical quantity describing the system in its nonequilibrium state with respect to the equilibrium one \cite{NABR19}. On the other hand, the random phase approximation (RPA) \cite{BP52, BP53, LP80} within the adiabatic-connection fluctuation-dissipation theorem \cite{LP75,LP77} predicted excellent equilibrium properties including bulk moduli of the copper-gold alloys \cite{NABR19}. The accurate formation energy of WB systems by these methods may be due to the nonlocality present in them \cite{ZKW14,NABR19}. On the contrary, we will later show that such nonlocality is not useful for SB systems; it improves the equilibrium volume but cannot correct the bulk moduli and formation energies.\\

Many of the earlier works focused mainly on the WB intermetallic alloys \cite{OWZ98,ZKW14,TLRKNV16,NABR19}, while a few others have included SB alloys \cite{IW18,ZSL07} using semilocal DFT approximations. In this work, we have explored a broad spectrum of binary intermetallic alloys from weakly- to strongly-bonded ones. The experimental formation energies of our test set range from 0.07 to 1.18 eV per atom. Most of the systems taken here are binary alloys that crystallize in the B2 (CsCl) phase. Table I presents the d-band metals with their electronic configurations. We have utilized one DFT approximation each from the different rungs of Perdew's Jacob ladder \cite{PS01}, except for LDA. The PBE-GGA (Perdew-Burke-Ernzerhof) \cite{PBE96}, the strongly constrained and appropriately normed (SCAN) \cite{SRP15}, the screened hybrid HSE06 (simply HSE by Heyd, Scuseria, and Ernzerhof) \cite{HSE03,HSE03-er}, and RPA were used as the DFT approximations. In this assessment, we aim to present a broader picture regarding the performance of various DFT approximations on these intermetallic alloys, which is missing from earlier works.\\

We will later classify the alloys into three different classes purely based on the performance of different DFT functionals for predicting formation energies, as shown in Figures~\ref{fig1} and~\ref{fig2}. We have a weakly-bonded (WB) region (Region I), where nonlocal HSE06 and RPA mostly perform better than semilocal PBE and SCAN. In the intermediate region (Region II), HSE06 and PBE work much better than others. Finally, there is the strongly-bonded (SB) region (Region III), where the nonlocal HSE06 and RPA severely fail to predict the accurate formation energy, while PBE-GGA outperforms them. Note that the classification performed here only refers to intermetallic alloys, and it is distinct from the classification adopted in Ref.~\cite{IW18} for alloys in general.

\begin{table}
	\resizebox{1.\columnwidth}{!}{%
	\begin{tabular}{c|cccccccccccc}
		\hline
		Element & Sc & Cu&Zn &Y &Rh &Pd &Ag &Cd &Hf &Os &Pt &Au \\ \hline
	    Configuration &3d$^1$4s$^2$ &3d$^{10}$4s$^1$ &3d$^{10}$4s$^2$ &4d$^{1}$5s$^2$ &4d$^{8}$5s$^1$ &4d$^{10}$ &4d$^{10}$5s$^1$ &4d$^{10}$5s$^2$ &5d$^{2}$6s$^2$ &5d$^{6}$6s$^2$ &5d$^{9}$6s$^1$ &5d$^{10}$6s$^1$ \\
	    \hline
		
		\end{tabular}
	}
\end{table}

\section{Computational details}
All DFT calculations were carried out using a projector augmented wave (PAW)  \cite{B94} method, as implemented in VASP  \cite{VASP} and GPAW  \cite{gpaw1,gpaw2,ase}. We performed spin-polarized semilocal and HSE06 calculations using VASP, while we used GPAW to perform the RPA calculations. The semilocal calculations were initialized with the magnetic moments per site of 1.5 - 3.5 $\mu_B$, which converged to the nonmagnetic ground-state during self-consistency. The total energy is converged with respect to plane-wave cutoff and k-mesh for all methods within 1-5 meV/atom. Besides, separate convergence tests for the EXX and correlation energies were performed for RPA. The detailed information about the plane-wave cutoff and Brillouin zone sampling are given in Supplementary Tables S1 and S2. Spin-unpolarized ground state PBE calculations were performed as inputs for the non-self-consistent RPA (for both EXX and correlation energies). We used a maximum cutoff of 350 eV to compute the response function. The correlation energies were computed as a function of the cutoff energy and extrapolated to infinity to get the RPA correlation energy as described in Refs.~\cite{HK08}. The gamma point (\textbf{q} $=$ 0) was skipped to avoid the possible divergent contribution and for smooth convergence with respect to k-mesh \cite{HSK10}, as required for metallic systems. We used the recommended PBE pseudopotentials (PP) modified to include the kinetic energy density for VASP calculations \cite{KJ99}, while 0.9.20000 data sets were utilized for GPAW calculations. The calculations include relativistic effects at the scalar level for each atom within the PAW PP.\\

Previously, it was showed that the spin-orbit coupling, zero-point vibrational energy, and the nonlocal vdW corrections have negligible effects on the formation energies of intermetallic alloys \cite{NABR19}. Therefore, we did not calculate those corrections in the present assessment. We have performed calculations for 7 volume points near the experimental equilibrium volume and fit the Birch-Murnaghan equation of state  \cite{B71} to evaluate the equilibrium properties. We have used the structures from Ref.~\cite{OKS96} and varied the lattice constants isotropically to generate structures with different volumes.

\section{Results and Discussions}
In this section, we discuss the results for ground-state equilibrium properties of various binary intermetallic alloys using DFT calculations. We present and discuss the equilibrium volumes and the bulk moduli in Supplementary Tables S2 and S3 separately. Here, we mainly focus on the performance of various DFT approximations in predicting formation energies. The formation energy is an important physical quantity in alloy theory as it governs the stability of that alloy. Suppose A$_x$B$_{1-x}$ is a binary alloy with constituent metals A and B. Then, the formation energy per atom $\Delta E_f$ can be computed as
\begin{equation}
 \Delta E_f(\text{A}_x\text{B}_{1-x}) = E(\text{A}_x\text{B}_{1-x}) - xE(\text{A}) - (1-x)E(B).
\end{equation}

\noindent where, $E(\text{A}_x\text{B}_{1-x})$, $E(\text{A})$, and $E(\text{B})$ are the total bulk energies per atom of an alloy $A_xB_{1-x}$, metal A, and metal B respectively. $x$ is the fractional weight of metal A in an alloy. A positive formation energy represents an  instability, while the negative value depicts the stability of an alloy against its constituents.\\

\begin{table*}
	\caption{Formation energy (eV) per atom of intermetallic alloys; the first column represents the alloys, while the second column shows the combination of the d-bands; CF is completely filled, PF is partially filled; All the compounds considered here crystallize in the B2 (CsCl) phase.}
	\begin{tabular}{lcrrrrc}
		\hline
		
		Alloys & Combination & \multicolumn{1}{l}{PBE} & \multicolumn{1}{l}{SCAN} & HSE06 & \multicolumn{1}{l}{RPA} & {Experiment} \\ \hline
		AgZn &4d(CF)-3d(CF) & -0.045 & -0.017 & -0.067 & -0.080 & -0.068 $\pm$ 0.002 \cite{HDHGK73} \\ 
		AgCd &4d(CF)-4d(CF) & -0.057 & -0.060 & -0.083 & -0.093 & -0.093 $\pm$ 0.002 \cite{HDHGK73} \\ 
		CuZn &3d(CF)-3d(CF) &-0.088 & -0.100 & -0.126 & -0.110 & -0.124 \cite{HDHGK73} \\ 
		CuPd & 3d(CF)-4d(CF) &-0.120 & -0.115 & {-0.193} & -0.168 & -0.131 $\pm$ 0.008 \cite{HDHGK73} \\ 
		AuCd & 5d(CF)-4d(CF) &-0.169 & -0.231 & {-0.196} & -0.244 & -0.196 \cite{HDHGK73} \\ 
		CuY &3d(CF)-4d(PF)  &-0.252 & -0.295 & -0.261 & \multicolumn{1}{l}{-0.354} & -0.200 $\pm$ 0.002 \cite{WK84,GK01}\\
		AuZn & 5d(CF)-3d(CF) &-0.211 & -0.260 & -0.252 & -0.290 & -0.235 $\pm$ 0.043  \cite{HDHGK73}\\ 
		ScAg & 3d(PF)-4d(CF) &-0.282 & -0.362 & -0.288 & -0.411 & -0.272 $\pm$ 0.017 \cite{FJK91,GK01}\\ 
		AgY & 4d(CF)-4d(PF) &-0.346 & -0.417 & -0.353 & \multicolumn{1}{l}{-0.467} & -0.278 $\pm$ 0.033 \cite{FJK91,GK01}\\ 
		HfOs & 5d(PF)-5d(PF) &-0.704 & -0.754 & {-0.923} & -0.667 & -0.482 $\pm$ 0.052 \cite{IW18,MG98}\footnote{It is mentioned in the reference that HfOs alloy sample was not completely pure. It had small amounts of Hf$_{54}$Os$_{17}$ and relatively important quantities of unreacted Os. Therefore, the true result should be more exothermic than -0.518 eV \cite{MG98}.} \\ 
		AuSc & 5d(CF)-3d(PF) &-0.812 & -1.063 & -0.854 & -1.045 & -0.789 $\pm$ 0.031 \cite{FJK91,GK01}\\ 
		RhY & 4d(PF)-4d(PF) &-0.841 & -0.939 & -0.979 & \multicolumn{1}{l}{-0.943} & -0.789 $\pm$ 0.035 \cite{SK93,GK01} \\ 
		ScPd & 3d(PF)-4d(CF) &-0.892 & -0.992 & -1.027 & -1.176 & -0.926 $\pm$ 0.023 \cite{SK91,GK01}\\ 
		ScRh & 3d(PF)-4d(PF) &-1.005 & -1.140 & -1.264 & -1.172 & -0.979 $\pm$ 0.016 \cite{SK91,GK01}\\ 
		PtSc &5d(PF)-3d(PF) & -1.216 & -1.455 & -1.382 & -1.440 & -1.086 $\pm$ 0.056 \cite{SK91,GK01}\\ 
		HfPt & 5d(PF)-5d(PF) &-0.972 & -1.195 & -1.073 & -1.058 & -1.178 $\pm$ 0.068 \cite{TK88,GK01}\\ \hline

	\end{tabular}
	\label{table1}
\end{table*}

\begin{figure}
	\includegraphics[scale=0.5]{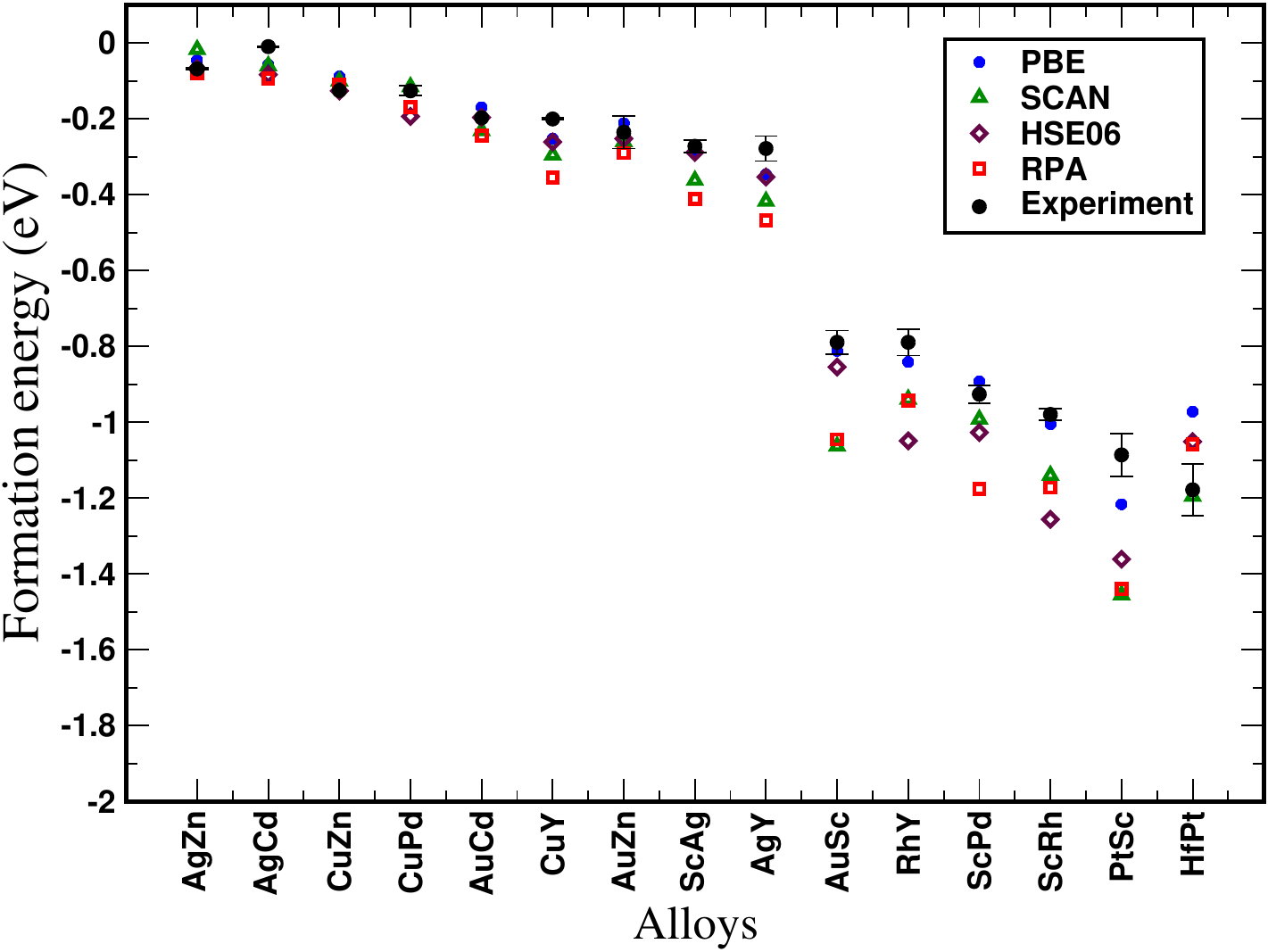}
	\caption{Computed formation energies compared with experimental values.}
	\label{fig1}
\end{figure}

We calculated the formation energies using various methods and tabulated  the data in Table~\ref{table1}. In Table I, we compare the results for an alloy crystallized in the B2 (CsCl) phase with available experimental data. The values are tabulated by an increasing magnitude of experimental formation energies from top to bottom. Figure~\ref{fig1} illustrates the graphical representation of the values in Table~\ref{table1}. Based on the observation, we classify the results into three distinct categories (or regions) depending on the filling of d-band in metals, as shown in Figure~\ref{fig2}. To understand the performance of functionals on formation energies, we plot the electronic density of states (DOS) of alloy and its constituents, because the structural stability of an alloy is largely dependent on the density of states \cite{P07}. We show that an accurate prediction of formation energy should be accompanied by an accurate prediction of the DOS.
\begin{enumerate}
	\item Completely-Filled / Completely-Filled (CF-CF)
	\item Completely-Filled / Partially-Filled (CF-PF)
	\item Partially-Filled / Partially-Filled (PF-PF)
\end{enumerate}

\subsection{(I) CF-CF combination}
\begin{figure}
	\centering
	\begin{subfigure}{0.5\textwidth}
		\includegraphics[scale=0.33]{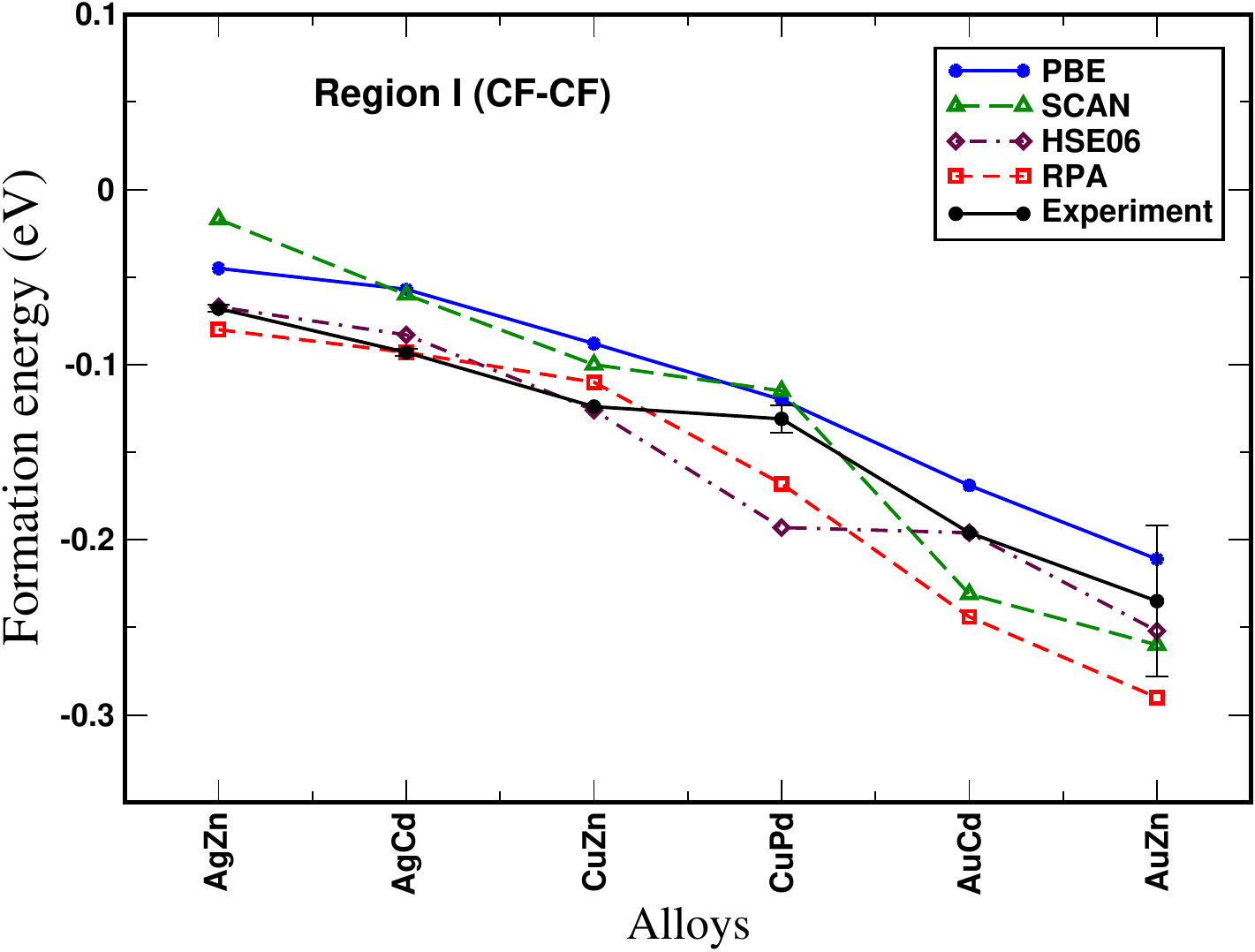}
		\caption{CF-CF combination}
		\label{fig:CF-CF1}
	\end{subfigure}
	\begin{subfigure}{0.5\textwidth}
		\includegraphics[scale=0.33]{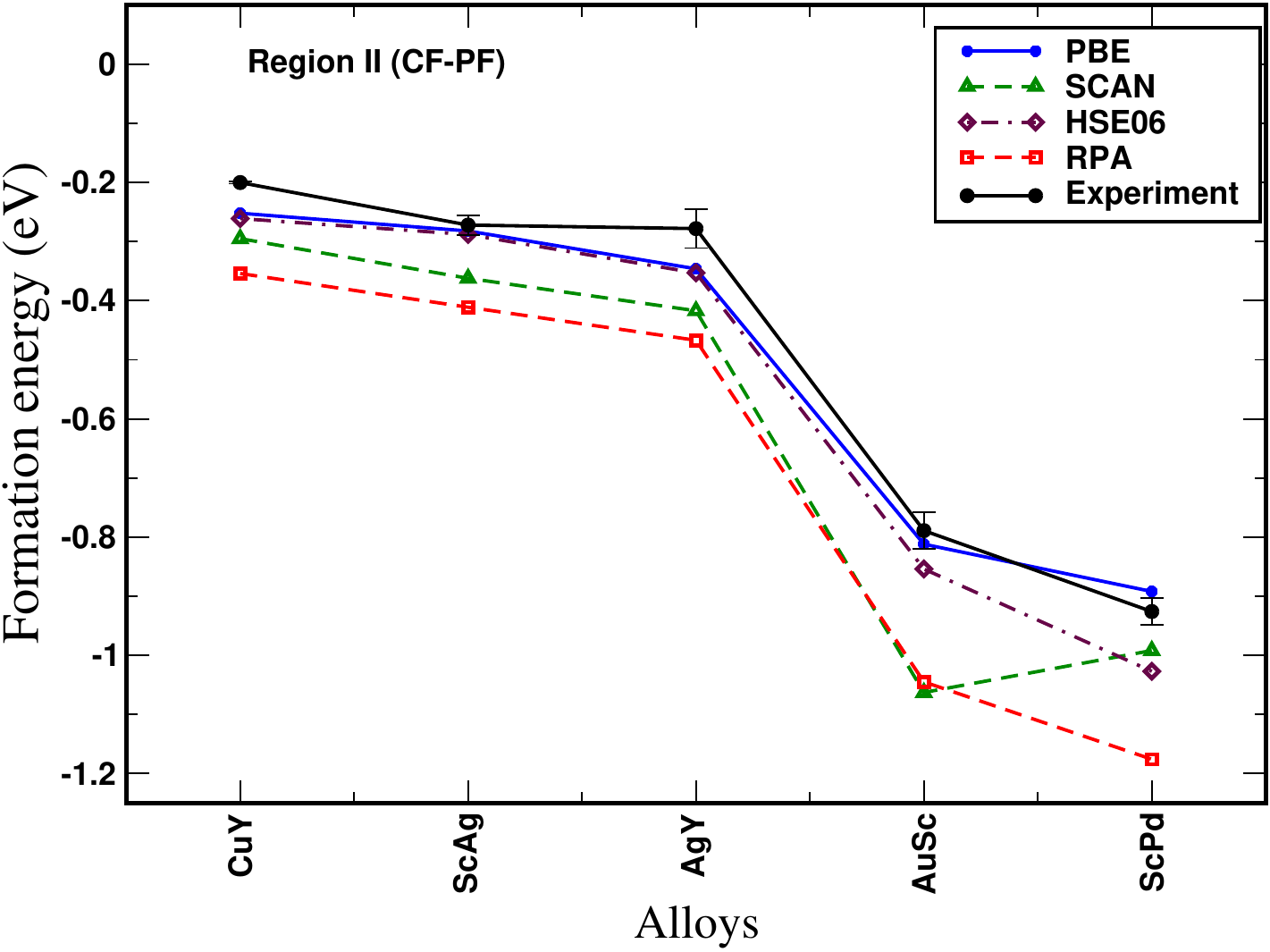}
		\caption{CF-PF combination}
		\label{fig:CF-PF1}
	\end{subfigure}%
	\begin{subfigure}{0.5\textwidth}
		\includegraphics[scale=0.33]{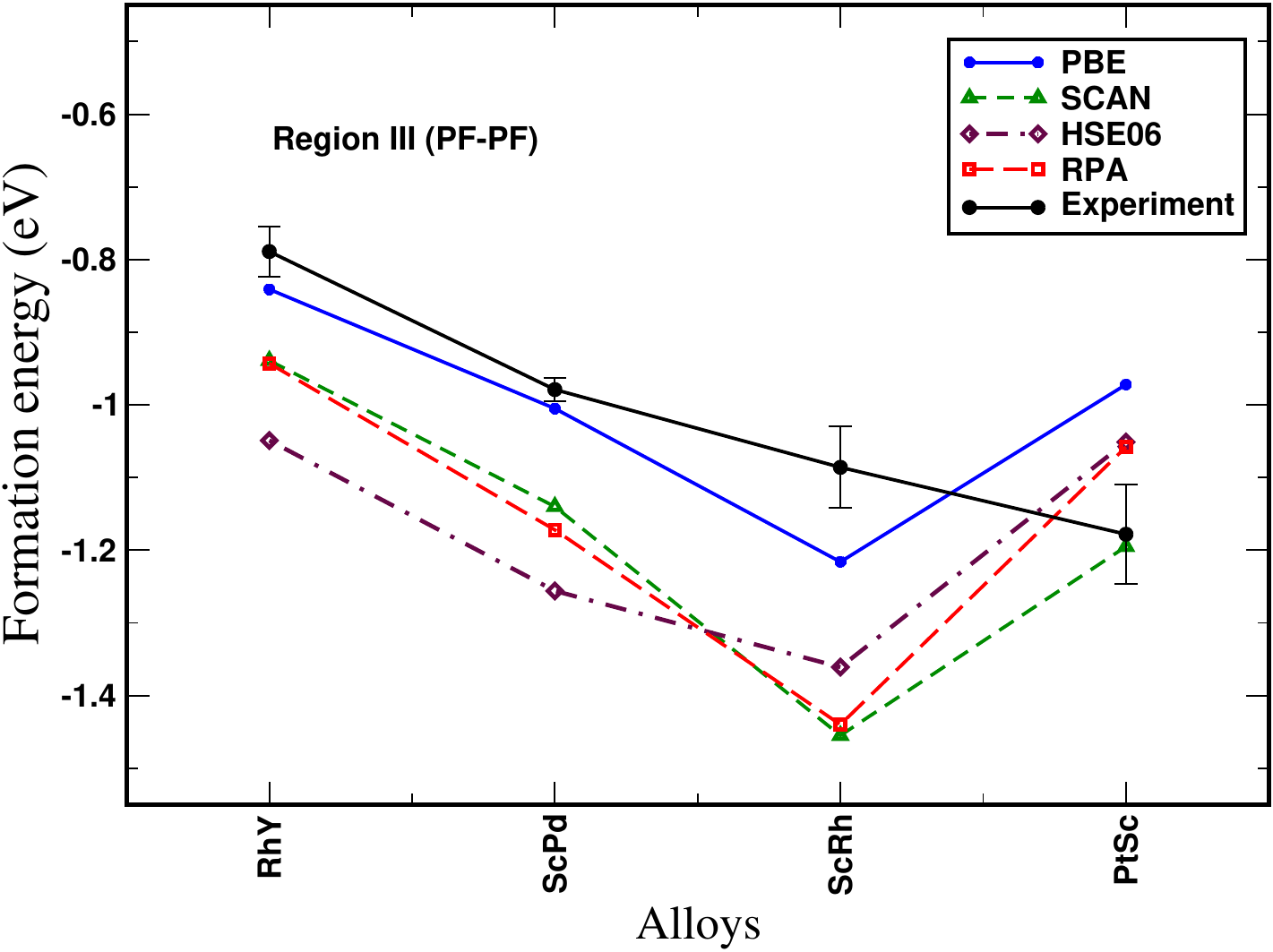}
		\caption{PF-PF combination}
		\label{fig:PF-PF1}
	\end{subfigure}%
	\caption{Formation energies with respect to three distinct classes.}
	\label{fig2}
\end{figure}
This set consists of intermetallic alloys with constituents bulk metals having completely-filled d bands such as AgZn, AgCd, CuZn, CuPd, AuCd, and AuZn (see Fig.~\ref{fig:CF-CF1}). These weakly-bonded alloys have lower experimental formation energies up to 200 meV, compared to other combinations. In this region, the PBE-GGA underestimates the formation energies as expected. The SCAN meta-GGA slightly improves upon PBE and yields mixed results, i.e., underestimates the formation energies for systems AgZn, AgCd, CuZn, and CuPd, while overestimates them in the cases of AuCd and AuZn. In general, SCAN overestimates the experimental energies below CuPd with a formation energy larger than or equal to 130 meV. On the other hand, the hybrid HSE06 consistently predicts accurate formation energies compared to the experiment, and in agreement with the previous calculations on alloys of CF-CF combination \cite{ZKW14}. Our results are also valid for structures other than the B2 phase, which is evident from Table~\ref{table2} and Ref.~\cite{ZKW14}. This indicates that the phase of the crystal has an insignificant role in the performance of DFT approximations when predicting formation energies of intermetallic alloys. To describe the results, we investigate the electronic properties of the alloy and its constituent bulk metals. \\

\begin{figure}
	\centering
	\begin{subfigure}{0.5\textwidth}
		\centering
		\includegraphics[scale=0.45]{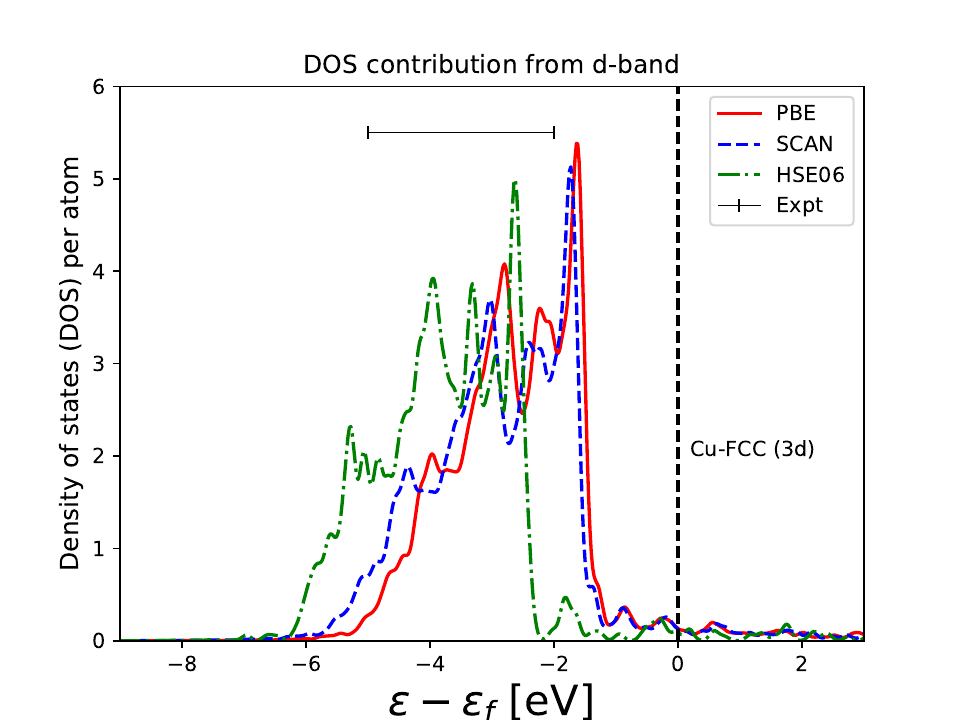}
		\caption{Copper \cite{AuCuAg}}
		\label{fig:cu}
	\end{subfigure}%
	\begin{subfigure}{0.5\textwidth}
		\centering
		\includegraphics[scale=0.45]{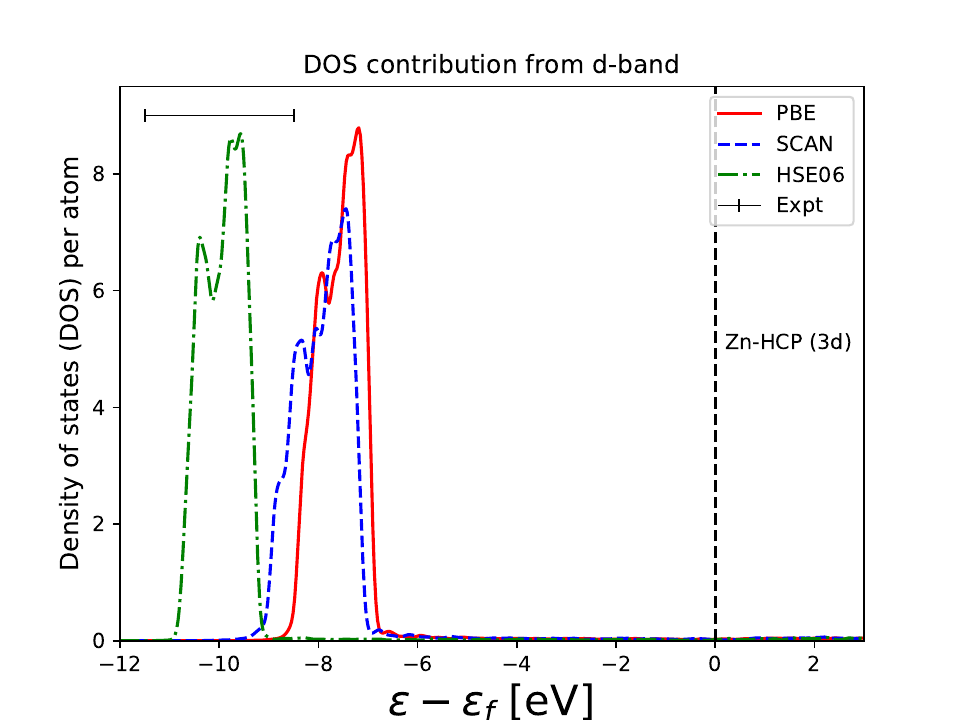}
		\caption{Zinc \cite{CuZn,AuZn,AgZn}}
		\label{fig:zn}
	\end{subfigure}
	\begin{subfigure}{0.5\textwidth}
		\centering
		\includegraphics[scale=0.45]{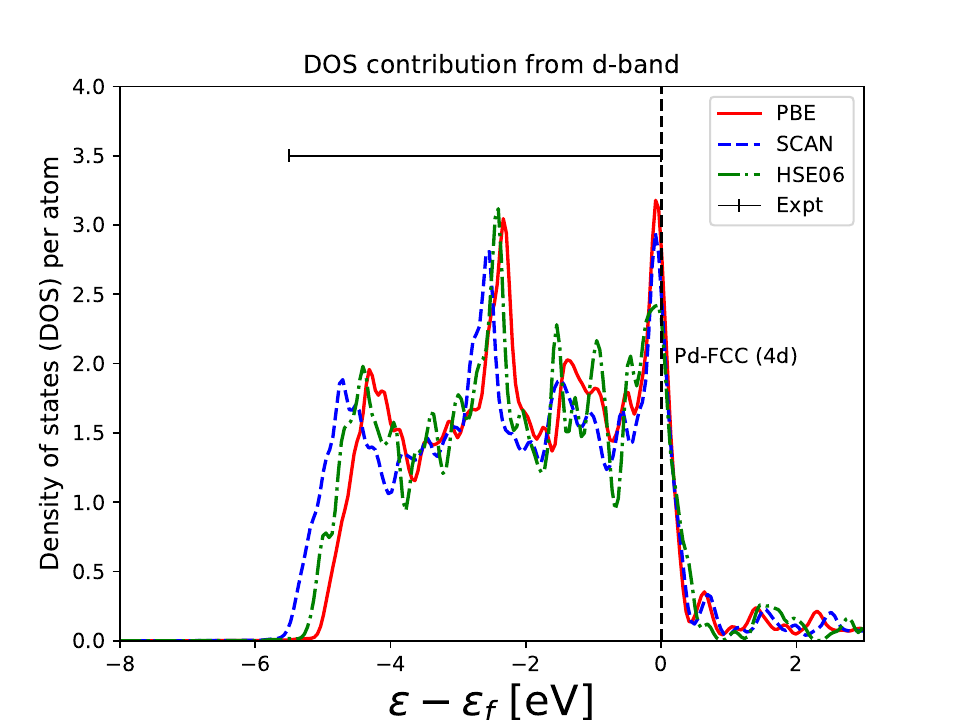}
		\caption{Palladium \cite{Pd,PdPt}}
		\label{fig:pd}
	\end{subfigure}%
	\begin{subfigure}{0.5\textwidth}
		\centering
		\includegraphics[scale=0.45]{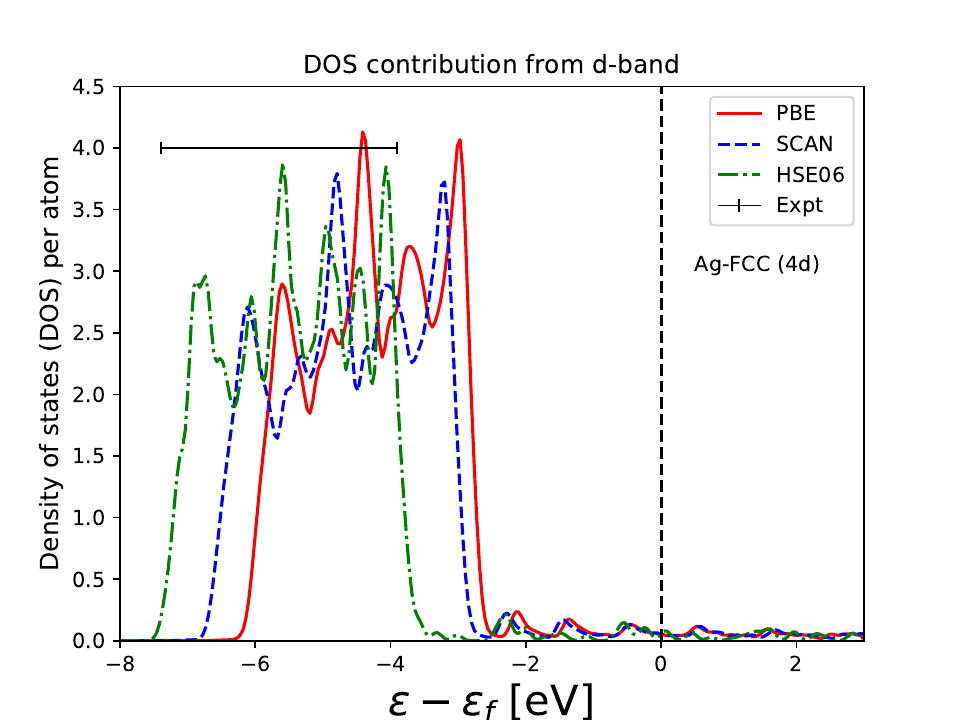}
		\caption{Silver \cite{AuCuAg,AgZn}}
		\label{fig:ag}
	\end{subfigure}
	\begin{subfigure}{0.5\textwidth}
		\centering
		\includegraphics[scale=0.45]{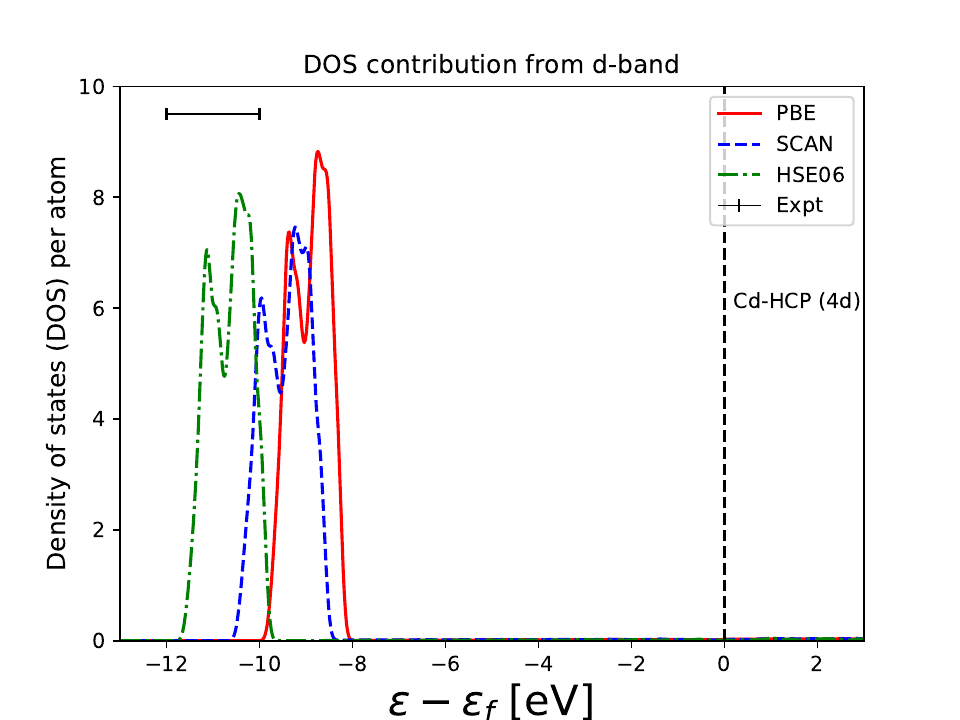}
		\caption{Cadmium \cite{AgCd,AgCd2}}
		\label{fig:cd}
	\end{subfigure}%
	\begin{subfigure}{0.5\textwidth}
		\centering
		\includegraphics[scale=0.45]{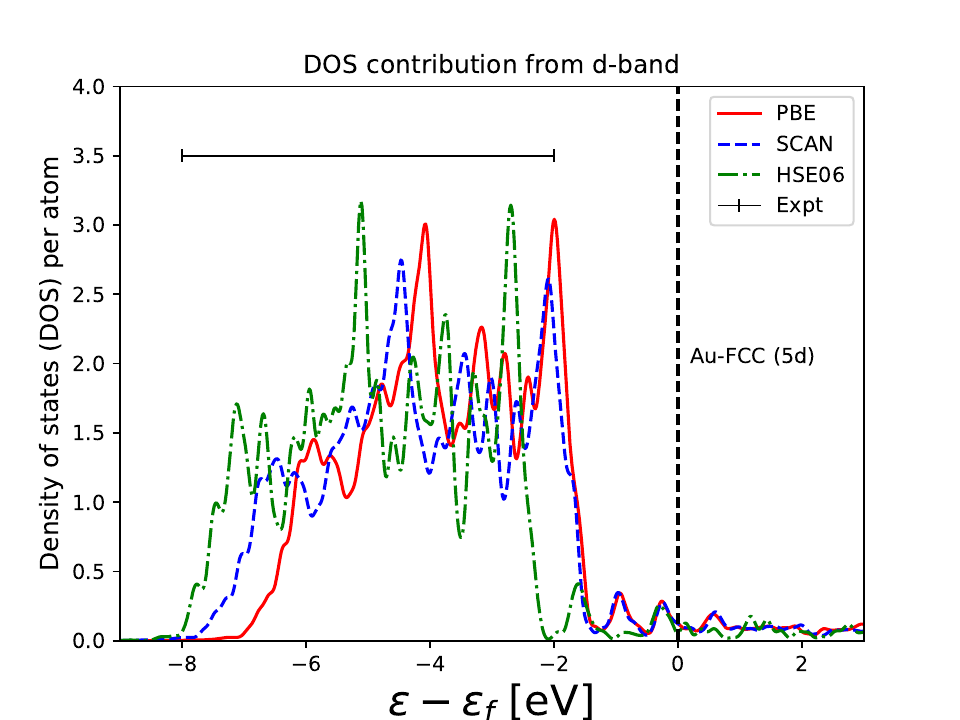}
		\caption{Gold \cite{AuCuAg,AuZn}}
		\label{fig:au}
	\end{subfigure}%
	\caption{The estimated valence d-band density of states with completely-filled d configuration compared with valence d-band ranges extracted from experimental X-ray photoemission spectra or ultraviolet photoemission spectra (denoted by horizontal solid line). References are given in sub-captions. $\epsilon_f$ is the Fermi-level.}
	\label{fig:CF}
\end{figure}

\begin{figure}
	\centering
	\begin{subfigure}{0.5\textwidth}
		\centering
		\includegraphics[scale=0.45]{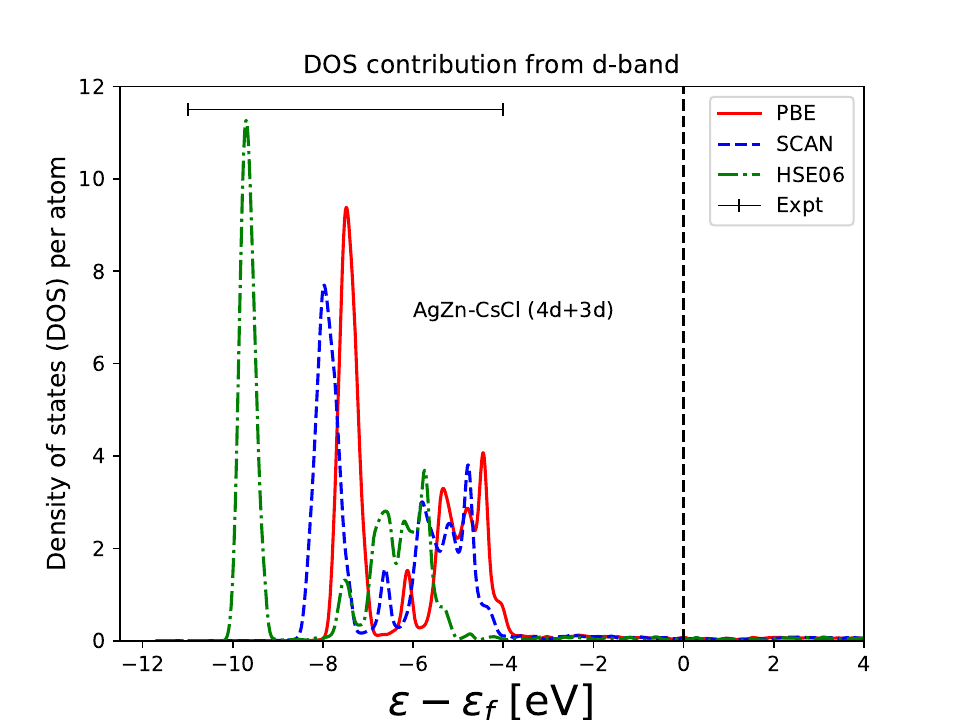}
		\caption{AgZn \cite{AgZn}}
		\label{fig:agzn}
	\end{subfigure}%
	\begin{subfigure}{0.5\textwidth}
		\centering
		\includegraphics[scale=0.45]{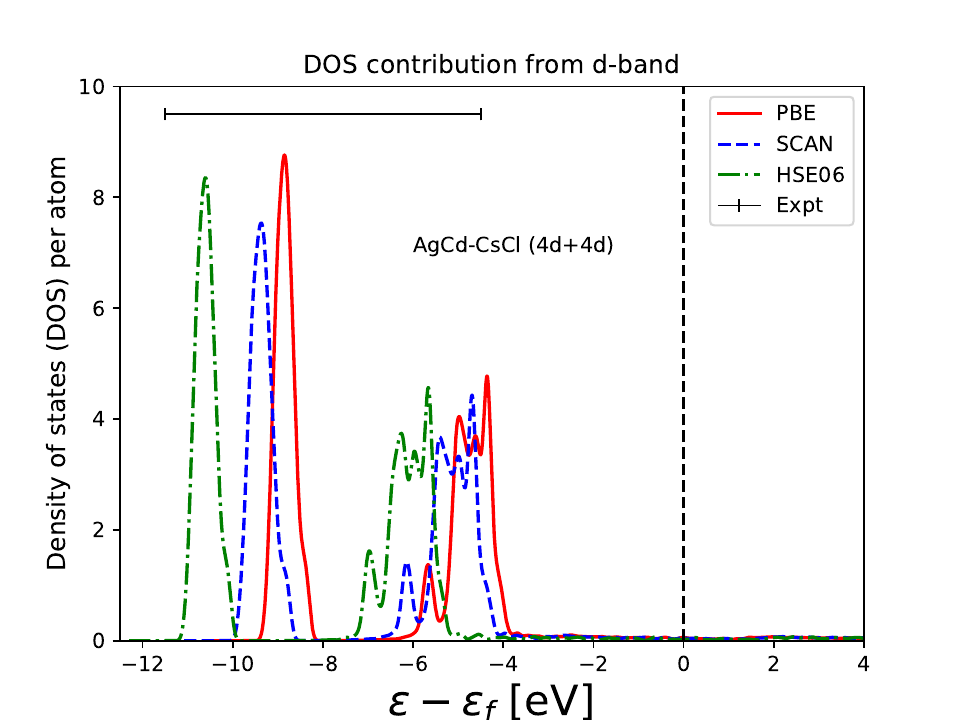}
		\caption{AgCd \cite{AgCd,AgCd2}.}
		\label{fig:agcd}
	\end{subfigure}
	\begin{subfigure}{0.5\textwidth}
		\centering
		\includegraphics[scale=0.45]{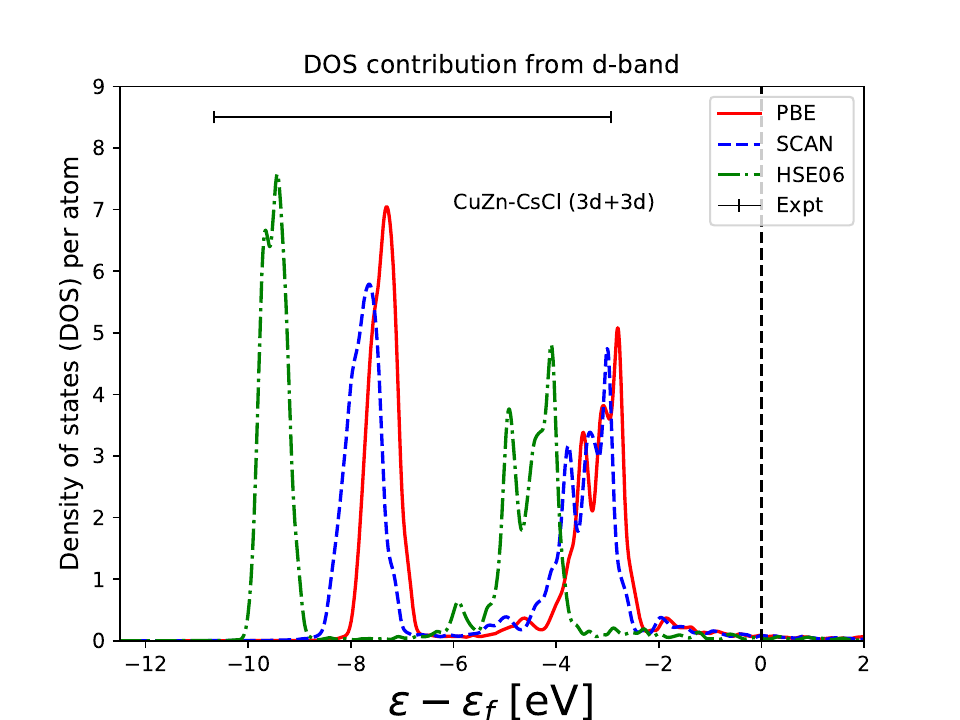}
		\caption{CuZn \cite{CuZn}}
		\label{fig:cuzn}
	\end{subfigure}%
	\begin{subfigure}{0.5\textwidth}
		\centering
		\includegraphics[scale=0.45]{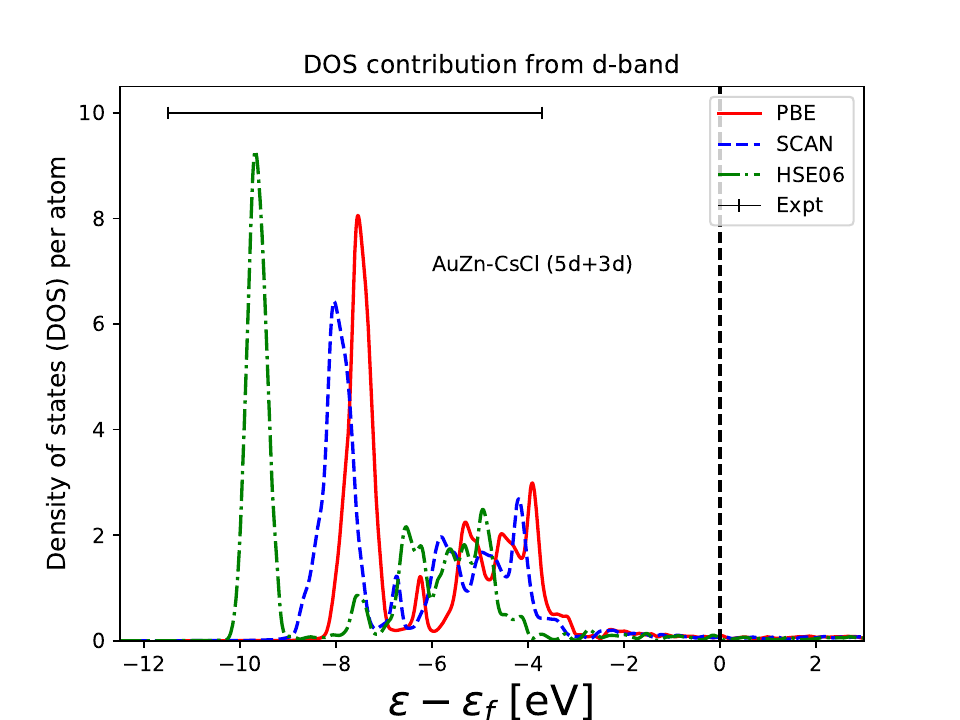}
		\caption{AuZn \cite{AuZn}}
		\label{fig:auzn}
	\end{subfigure}%
	\caption{The estimated valence d-band density of states of alloys with metals having completely-filled/completely-filled d-band configuration compared with valence d-band ranges extracted from experimental X-ray photoemission spectra or ultraviolet photoemission spectra (denoted by horizontal solid line). References are given in sub-captions. $\epsilon_f$ is the Fermi-level.}
	\label{fig:CF-CF}
\end{figure}

We computed valence-state electronic partial density of states (PDOS) for metals having completely-filled (CF) d-bands and alloys of CF-CF combination and present them in Figures~\ref{fig:CF} and \ref{fig:CF-CF}, respectively. The contribution from s- and p- bands near the Fermi level is negligible (not shown here) compared to that of d-band. Therefore, we only show the d-band contribution of the density of states (DOS) and compare it with the experimental PDOS of occupied states. Experimentally, one can obtain information about the valence d-band (or PDOS) from X-ray (XPS) or ultraviolet (UPS) photoemission spectra \cite{AuCuAg,CuZn,AgZn,AuZn,AgCd,ScPdScAg,Pd,PdPt,RhAg,Hf,Y,ScRh3} (see section ``Experimental data for valence d-band" in Supporting Information). \\

In Fig.~\ref{fig:CF}, we compare the calculated d-band contribution to DOS (PDOS) with experimental d-band ranges extracted from photoemission spectra (Supplementary Table S7). We present the PDOS for completely-filled metals such as copper, zinc, palladium, silver, cadmium, and gold with their phase and valence d-band. The negligible density of states at the Fermi level indicates the complete filling of valence bands (3d in Cu and Zn, 4d in Ag and Cd, and 5d in Au). Besides, Zn and Cd have even lesser PDOS (more than two factors in magnitude than other completely-filled d-band metals) at the Fermi level due to the filling of valence s-band as well. Copper has the 3d band centered around its binding energy ($E_b$) of 3.0$-$3.5 eV with a d-band width (or range) of $\sim$ 3 eV \cite{AuCuAg,CuZn}. Furthermore, Zinc and Cadmium have similar PDOS with localized valence 3d and 4d bands respectively centered around the binding energies $\sim$10 eV \cite{CuZn,AuZn,AgZn} and $\sim$11 eV \cite{AgCd,AgCd2} below the Fermi level with width $\sim$ 1.5$-$2.0 eV. On the other hand, Pd, Ag, and Au have d-band ranges 0$-$5.5 eV (4d) \cite{Pd,PdPt}, 3.9$-$7.4 eV (4d) \cite{AgZn}, and 2$-$8 eV (5d) \cite{AuCuAg,AuZn} below the Fermi level respectively. All DFT approximations agree with each other regarding the shape and width of the d-band in the PDOS plot. However, there is a discrepancy in the d-band center ($E_b$) among them. The PBE-GGA underestimates the binding energy (d-band center) of these metals by $\sim$ 1$-$2 eV (maximum for Cd and Zn), whereas SCAN provides a negligible improvement of $\sim$ 0$-$0.5 eV upon PBE. On the contrary, the hybrid HSE06 considerably improves on PBE and SCAN by increasing the binding energy (blue-shifted towards the experimental d-band range) and provides accurate results compared to experiments.\\

 Figure~\ref{fig:CF-CF} depicts the partial density of states of alloys AgZn (Fig.~\ref{fig:agzn}), AgCd (Fig.~\ref{fig:agcd}), CuZn (Fig.~\ref{fig:cuzn}), and AuZn (Fig.~\ref{fig:auzn}). In AgZn, the binding energy of Ag's 4d band increases by $\sim$ 1 eV (onset shifts from 3.0 to 4.0 eV in PBE and SCAN, and 4.0 to 5.0 eV in HSE06) and the width also shrinks by $\sim$ 1 eV compared to pure Ag's 4d band. On the other hand, zinc's 3d band is still localized around almost the same binding energy (small decrement though) as that of the pure Zn, with a negligible decrease in d-band width. These observations are consistent with the experimental results that the 4d band of Ag in AgZn decreases by 1 eV and the width shrinks by 0.7 eV with a negligible effect on the zinc's 3d band compared to its pure constituents \cite{AgZn}. Similar results can be obtained for AgCd, CuZn, and AuZn. Ag's 4d band in AgCd, Cu's 3d band in CuZn, and Au's 5d band in AuZn behave similarly to that of the Ag's 4d band in AgZn. Also, Zn and Cd in these compounds behave likewise as that of Zn in AgZn. The sharp decrease in d-band widths of Cu, Ag, and Au in alloys can be attributed to the dilution (it increases the distance between two CF metal nearest neighbors which decreases its overlap with them, thereby giving localized and bound state) of these metals in alloys in presence of more localized 3d and 4d band of Zn and Cd respectively \cite{AgZn,AgCd,CuZn,AuZn}. Also, the increase in binding energy in one of the metals and decrease in the binding energy in the other indicates some charge transfer between the constituents \cite{CuZn,AuZn}. Qualitatively, all DFT functionals PBE, SCAN, and HSE06 yield similar results in terms of the change in PDOS of alloys with respect to its constituents. However, only the nonlocal HSE06 provides an accurate binding energy in the case of both alloys and constituents, thereby predicting accurate formation energies. Also, the nonlocal RPA shows similar or better accuracy than that of semilocal functionals in predicting formation energies of CF-CF alloys. We will discuss the RPA results in detail later in a separate section ``Failure of RPA and beyond RPA correction".      

\begin{table}
	\caption{Formation energies (eV) per atom of alloys other than the B2 phase (CsCl, Pm-3m); CF is completely filled, PF is partially filled.}
	\begin{tabular}{lcccccc}
		\hline
			Alloys & Phase &combination &\multicolumn{1}{l}{PBE} & \multicolumn{1}{l}{SCAN} & HSE06 & {Expt} \\ \hline
%		VPt$_2$ & MoPt$_2$ (Immm) & \multicolumn{1}{l}{-0.558} & \multicolumn{1}{l}{-0.733} & -- & -0.386 $\pm$ 0.026 \cite{GK94} \\ 
		AuTi & CuTi (P4/nmm) &5d(CF)-3d(PF) & -0.442 & -0.630 & -0.430 & -0.458 $\pm$ 0.015 \cite{FK92} \\ 
%		Au$_2$Ti & MoSi$_2$ (I4/mmm) & -0.445 & -0.651 & -- & -0.476 $\pm$ 0.021 \cite{FK92}\\ 
		ScRh$_3$ & AuCu3 (Pm-3m)&3d(PF)-4d(PF) &-0.610 & -0.762 & -0.770 & -0.536 $\pm$ 0.015 \cite{SK91}\\ 
		YPd$_3$ & AuCu3 (Pm-3m)& 4d(PF)-4d(CF) &-0.867 & -0.890 & -1.041 & -0.819 $\pm$ 0.067 \cite{SK93} \\
		ScPt$_3$ & AuCu3 (Pm-3m) & 3d(PF)-5d(PF)&-1.042 & -1.263 & -1.173  & -0.980 $\pm$ 0.021 \cite{GK95}\\  
		
		\hline
	\end{tabular}
	\label{table2}
\end{table}

\subsection{(II) CF-PF combination}
In this section, we discuss the results for alloys with completely-filled / partially-filled (CF-PF) d-band combinations such as CuY, ScAg, AgY, AuSc, and ScPd having a B2 phase, shown as in Fig.~\ref{fig:CF-PF1}. We also present a few other alloys (AuTi and YPd$_3$) with a different structure than B2 in Table~\ref{table2}. The formation energies predicted by PBE-GGA agree well with the experiment, whereas the hybrid HSE06 concurs with PBE for lower formation energies (200$-$500 meV), while differs significantly from PBE and experimental results in the case of higher formation energies (e.g., YPd$_3$ and ScPd). On the contrary, SCAN seriously overestimates the formation energies in this region with a decrease in its deviation for alloys with increasing formation energy (YPd$_3$ and ScPd). The nonlocal RPA consistently overestimates the formation energies of alloys with results closer to SCAN at the best-case scenario.\\

\begin{figure}
	\centering
	\begin{subfigure}{0.5\textwidth}
		\centering
		\includegraphics[scale=0.45]{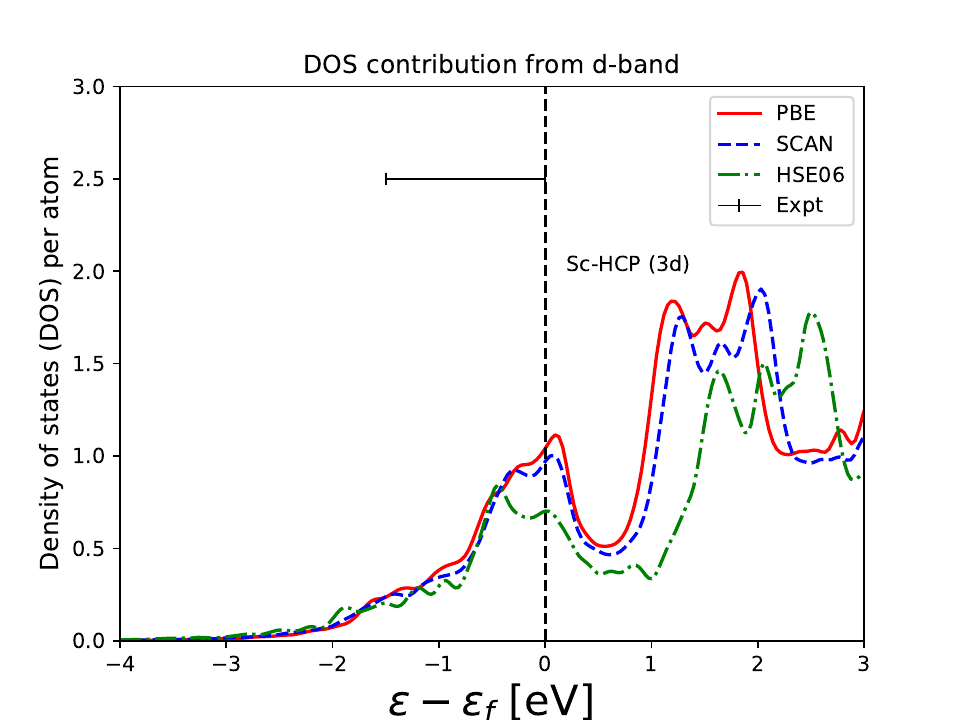}
		\caption{Scandium \cite{ScPdScAg}}
		\label{fig:sc}
	\end{subfigure}%
		\begin{subfigure}{0.5\textwidth}
			\centering
			\includegraphics[scale=0.45]{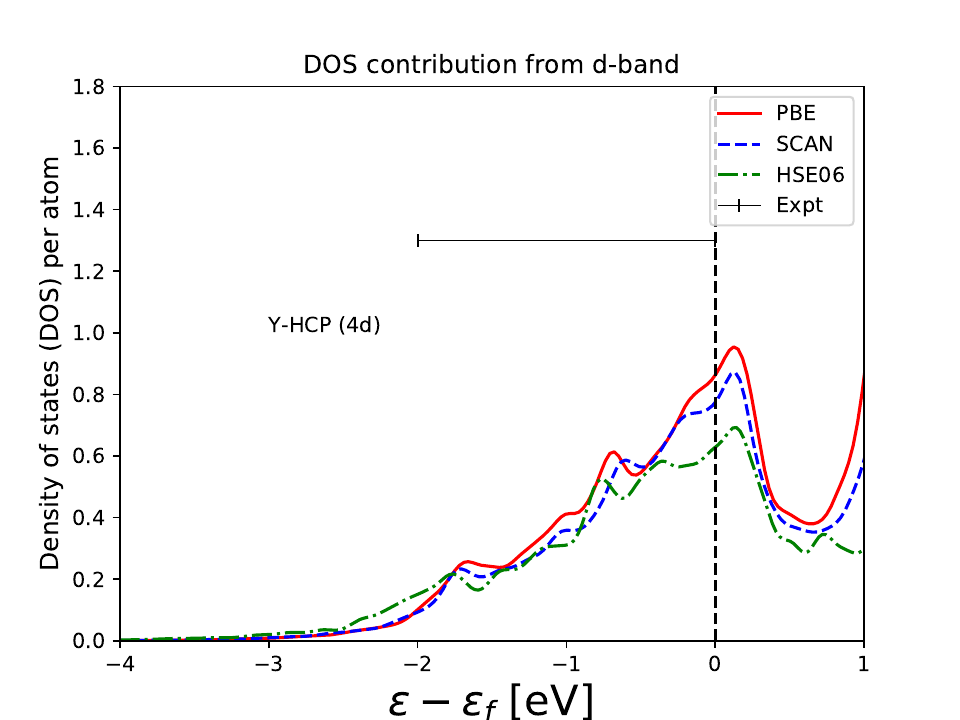}
			\caption{Yttrium \cite{Y}}
			\label{fig:y}
		\end{subfigure}
	\begin{subfigure}{0.5\textwidth}
		\centering
		\includegraphics[scale=0.45]{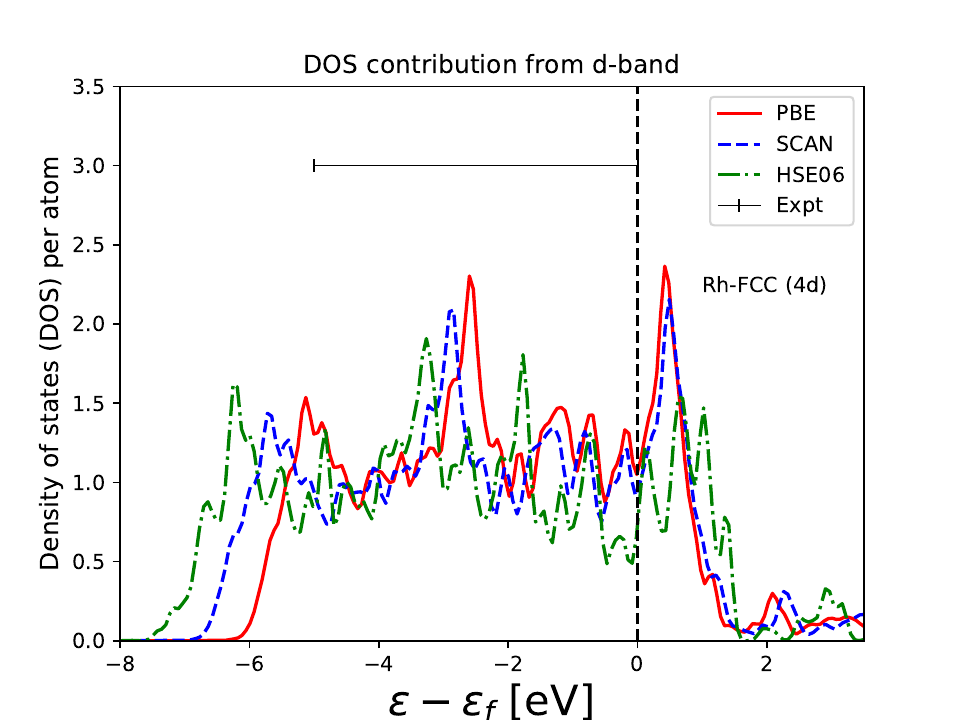}
		\caption{Rhodium \cite{RhAg,Rh}}
		\label{fig:rh}
		\end{subfigure}%
		\begin{subfigure}{0.5\textwidth}
			\centering
			\includegraphics[scale=0.45]{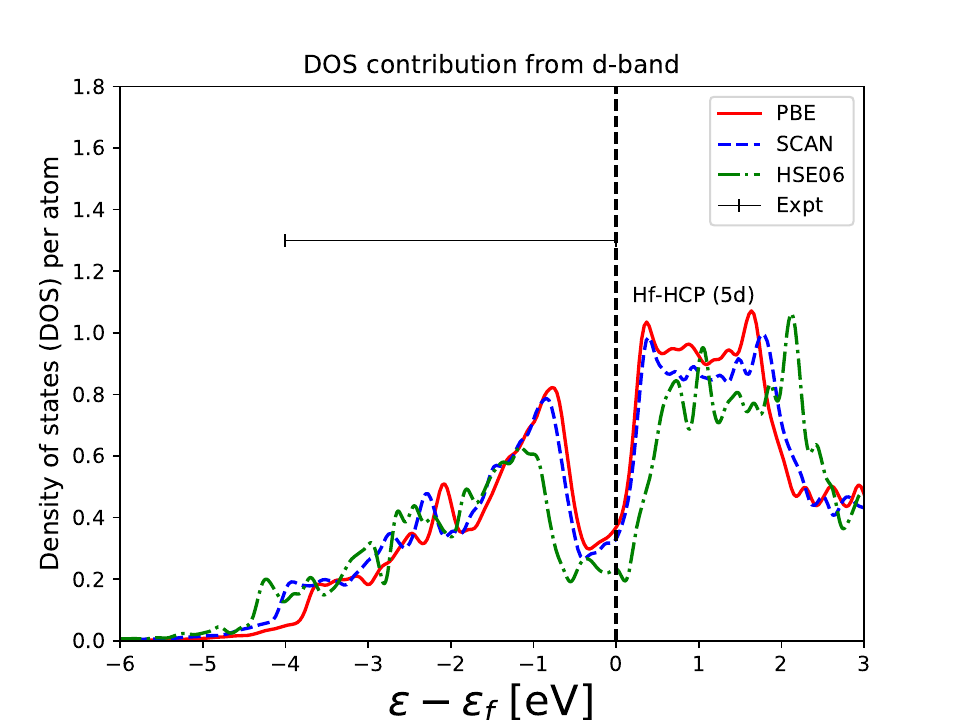}
			\caption{Hafnium \cite{Hf}}
			\label{fig:hf}
		\end{subfigure}
			\begin{subfigure}{0.5\textwidth}
				\centering
				\includegraphics[scale=0.45]{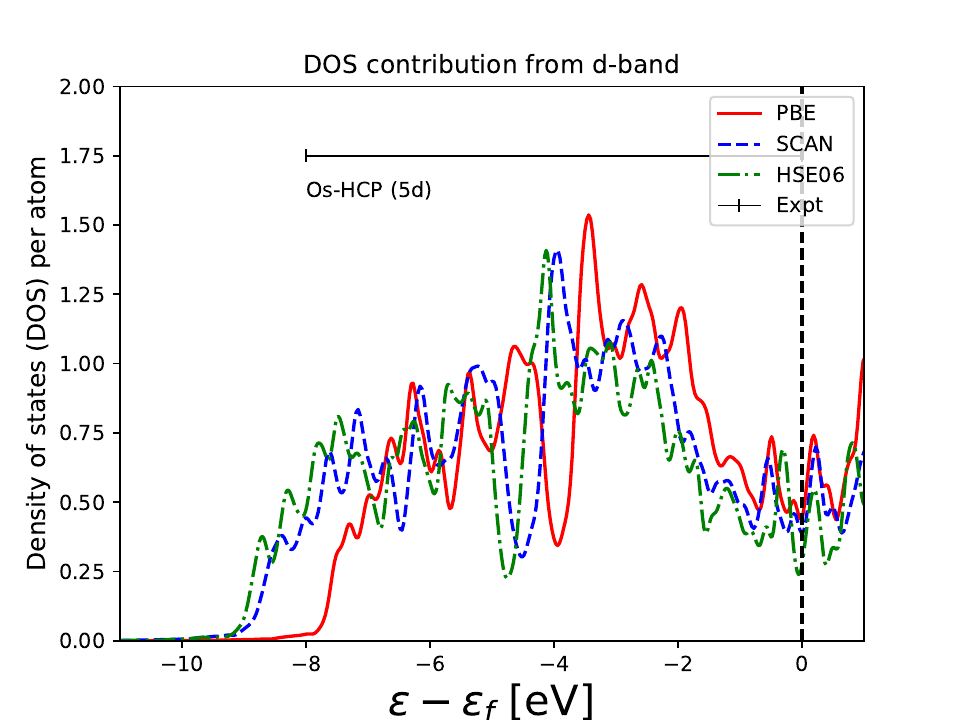}
				\caption{Osmium \cite{Rh}}
				\label{fig:os}
			\end{subfigure}%
		\begin{subfigure}{0.5\textwidth}
		\centering
		\includegraphics[scale=0.45]{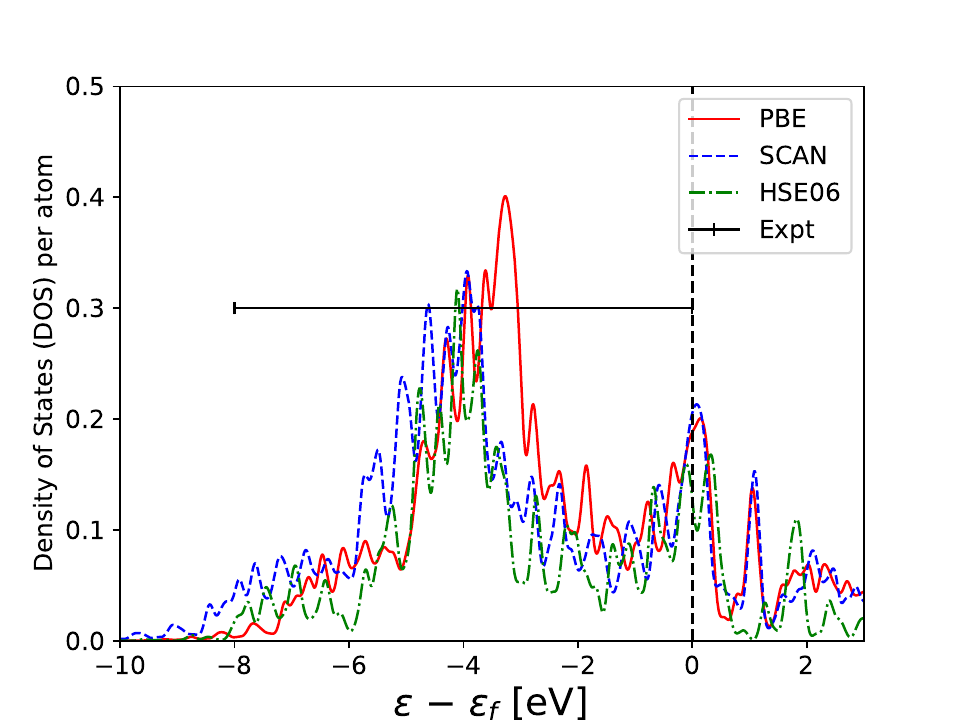}
		\caption{Platinum \cite{PdPt}}
		\label{fig:pt}
	\end{subfigure}
	
	\caption{The estimated valence d-band density of states with partially-filled d configuration compared with valence d-band ranges extracted from experimental X-ray photoemission spectra or ultraviolet photoemission spectra (denoted by horizontal solid line). References are given in sub-captions. $\epsilon_f$ is the Fermi-level.}
	\label{fig:PF}
\end{figure}

Partial density of states (PDOS) of valence d-band of partially-filled (PF) metals scandium (3d), yttrium (4d), rhodium (4d), hafnium (5d), osmium (5d), and platinum (5d) are shown in Figure~\ref{fig:PF}. Unlike the completely-filled metals, there is a significantly large density of states at the Fermi level. Experimentally, both scandium's 3d and yttrium's 4d band is localized near the Fermi level ($\sim$ 0.2 eV below for Sc) with the d-band widths of $\sim$ 1.5 eV and 2.0 eV respectively \cite{ScPdScAg,Y}. Also, rhodium's 4d band is concentrated at $\sim$ 1.3$-$1.5 eV below the Fermi-level and has a d-band range of $\sim$ 4.5$-$5.0 eV \cite{Rh,RhAg}. The 5d bands of both Hf and Pt have similar localization as that of the 4d band of Rh with centroids around $\sim$ 0.9 eV and $\sim$ 1.6 eV respectively, with d-band widths of $\sim$ 4 eV and $\sim$ 8 eV \cite{Hf,Rh,PdPt} (Pt's 5d-band has two peaks, and it is the first peak). Similary, osmium's 5d band has a peak at $\sim$ 3.0 eV and a d-band width of $\sim$ 8.0 eV \cite{Rh}. For these alloys, the PDOS at the Fermi-level decreases as PBE $>$ SCAN $>$ HSE06, except for platinum. Also, the d-band range is blue-shifted away from the experimental valence d-band range in the order of PBE $<$ SCAN $<$ HSE06 with PBE being the closest. However, such a shift is noticeable only in the Rh's 4d and Os's 5d bands. Unfortunately, it is difficult to pinpoint the peak or a d-band centroid of DFT calculated PDOS for Rh's 4d, Os's 5d, and Pt's 5d bands, therefore, we could not compare it directly with the experimental values.\\

\begin{figure}
\centering
	\begin{subfigure}{0.5\textwidth}
		\centering
		\includegraphics[scale=0.45]{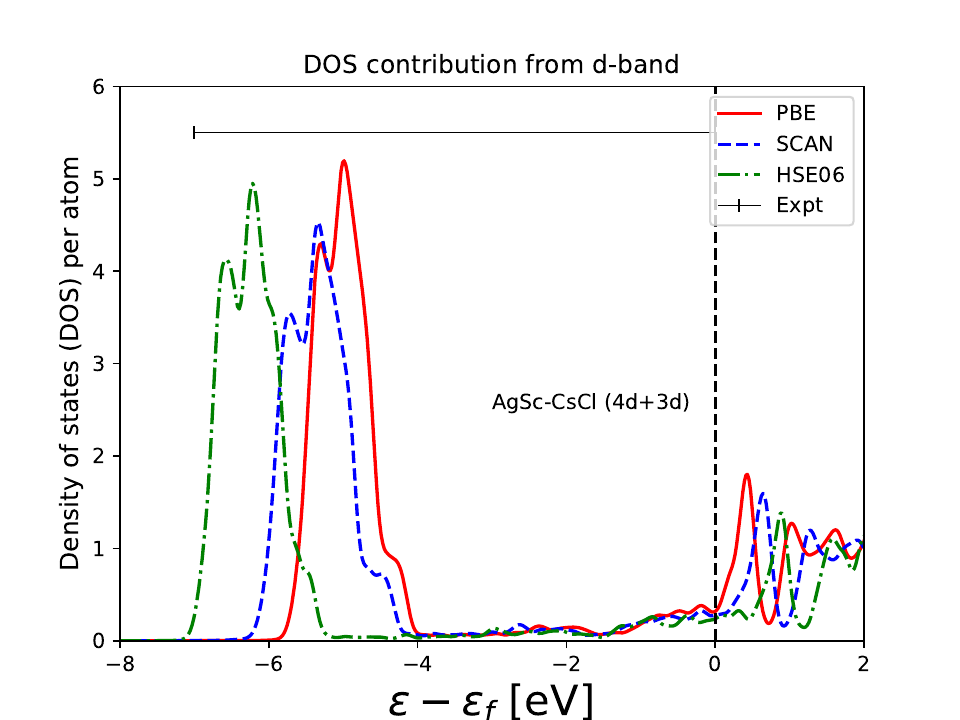}
		\caption{AgSc \cite{ScPdScAg}}
		\label{fig:agsc}
	\end{subfigure}
	\begin{subfigure}{0.5\textwidth}
		\centering
		\includegraphics[scale=0.45]{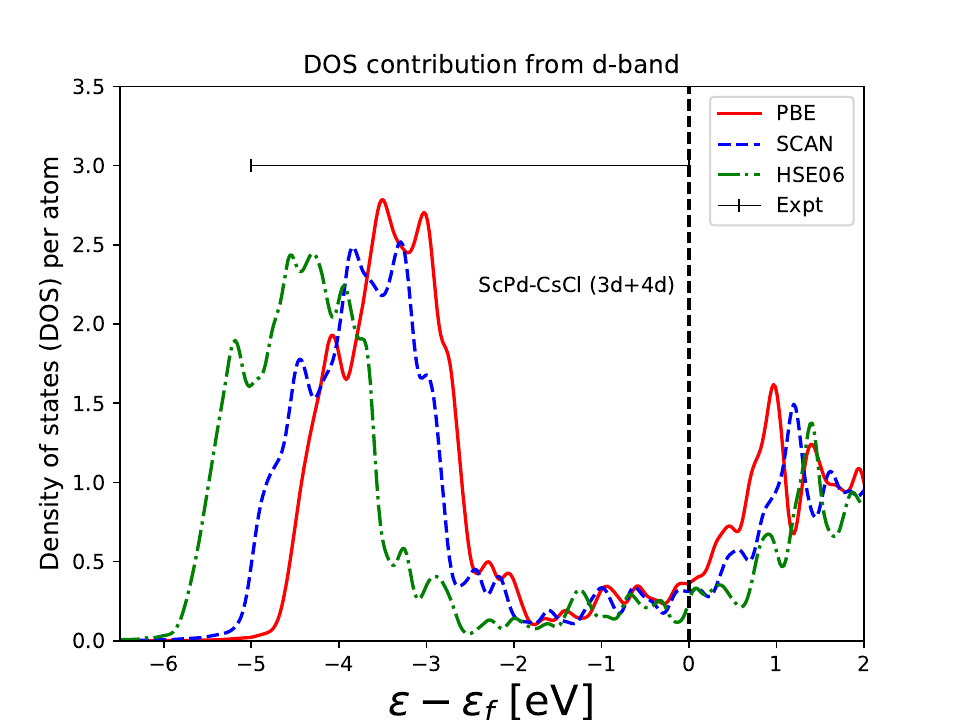}
		\caption{ScPd \cite{ScPdScAg}}
		\label{fig:scpd}
	\end{subfigure}%
	 	\begin{subfigure}{0.5\textwidth}
	 		\centering
	 		\includegraphics[scale=0.45]{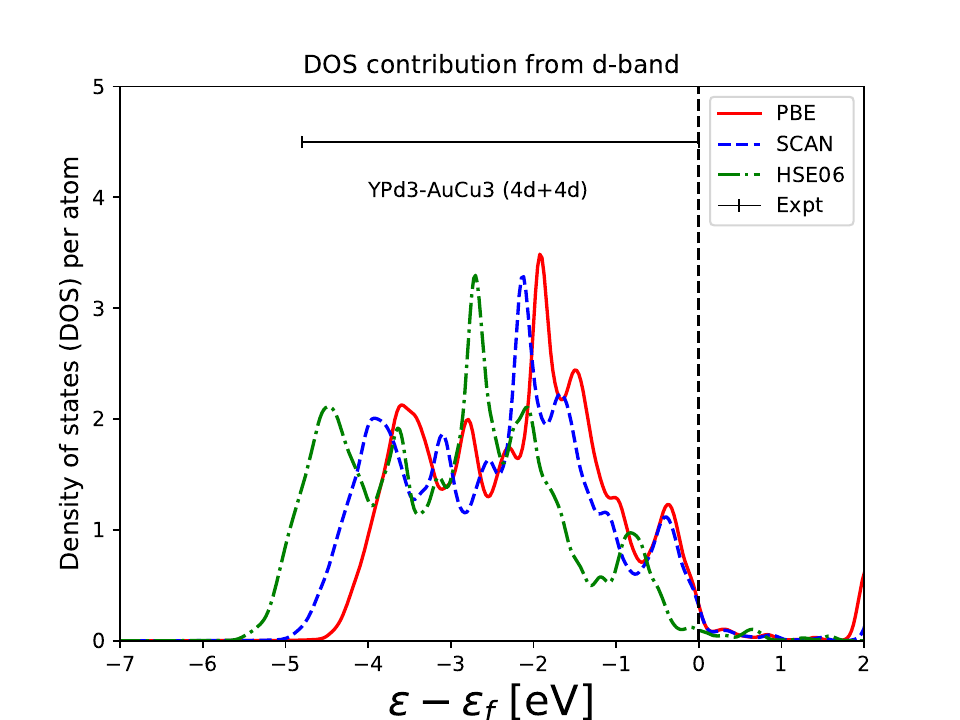}
	 		\caption{YPd$_3$ \cite{YPd3,Pd}}
	 		\label{fig:ypd3}
	 	\end{subfigure}%
	\caption{The estimated valence d-band density of states of alloys with metals having completely-filled / partially-filled d-band configuration compared with valence d-band ranges extracted from experimental X-ray photoemission spectra or ultraviolet photoemission spectra (denoted by horizontal solid line). References are given in sub-captions. $\epsilon_f$ is the Fermi-level.}
	\label{fig:cf-pf2}
\end{figure}

Figure~\ref{fig:cf-pf2} shows PDOS results for alloys AgSc, ScPd, and YPd$_3$. In both ScPd and ScAg, the large density of states at the Fermi-level mainly consists of the partially-filled 3d band of scandium, and its tail is slightly stretched towards larger binding energies compared to its pure counterpart (Sc in ScAg is more stretched than Sc in ScPd) \cite{ScPdScAg}. Our PDOS calculated by all DFT functionals qualitatively agree with this experimental observation (see Figures~\ref{fig:agsc} and \ref{fig:scpd}). On the other hand, the completely-filled d-band of metals Ag and Pd should be red-shifted towards a lower binding energy with a decrease in its width due to similar reasons of charge-transfer and dilution as in the case of CF-CF alloys. It is the part where our DFT calculated PDOS differs from the experiment. As expected, PBE underestimates the d-band centroid of the Ag's and Pd's 4d bands, while SCAN slightly blue-shifts them. HSE06 also raises the binding energy, however, the shifted 4d band of Ag agrees with the experiment, but is overestimated in the case of Pd's 4d band. This could be the reason that the formation energy predicted by HSE06 is accurate for ScAg, but overestimates the ScPd formation energy. The calculated electronic PDOS of YPd$_3$ has nonseparable 4d bands of yttrium and palladium, similar to the experimental photoemission spectrum \cite{Pd}. Nevertheless, it has different centroids and d-band ranges for different DFT methods. Similar to ScPd, the inaccuracy of HSE06 in the prediction of formation energy of YPd$_3$ is due to an incorrect prediction of the electronic PDOS of an alloy. Conversely, PBE and SCAN provide a reliable estimate of formation energies for both ScPd (Table~\ref{table1}), and YPd$_3$ (Table~\ref{table2}), because these functionals predict the correct electronic properties of both alloys and the bulk elements simultaneously. \\

\subsection{(III) PF-PF combination}
Previously, we have observed that the inclusion of partially-filled d-band metals in alloys diminish the accuracy of the nonlocal density functionals, while the PBE-GGA consistently outperforms them. The SCAN meta-GGA indeed improves the PBE calculated electronic properties of many alloys and elemental bulks, but does not possess similar accuracy in formation energies as that of PBE. Here, we explore the alloys with both partially-filled d-band metals such as HfOs, RhY, ScRh, PtSc, HfPt, ScRh$_3$, and ScPt$_3$. In the PF-PF combination of alloys, both nonlocal functionals, HSE06 and RPA, severely overestimate the formation energies. On the other hand, semilocal functionals perform well with PBE performing much better than the SCAN except for HfPt. Our DFT results show a significant error (overestimation) in the case of HfOs, even for PBE. We suspect that there are some uncertainties in the experiment, as a few traces of Hf$_{54}$Os$_{17}$ and unreacted Os are also present in the sample \cite{MG98}. \\

\begin{figure}
	\centering
		\begin{subfigure}{0.5\textwidth}
			\centering
			\includegraphics[scale=0.45]{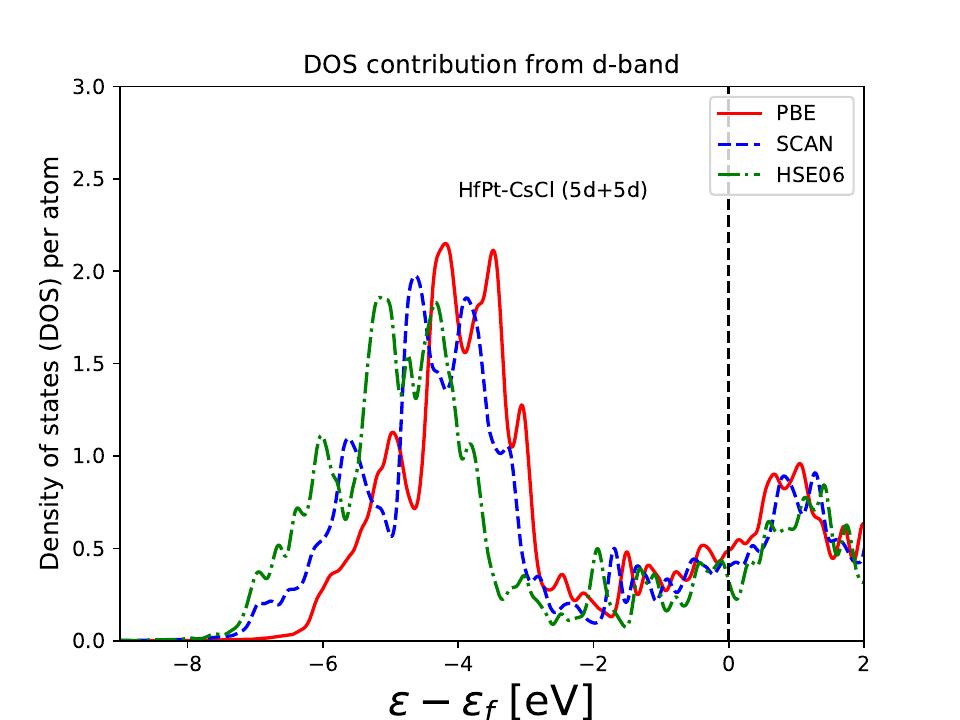}
			\caption{HfPt}
			\label{fig:hfpt}
		\end{subfigure}%
   	\begin{subfigure}{0.5\textwidth}
   		\centering
   		\includegraphics[scale=0.45]{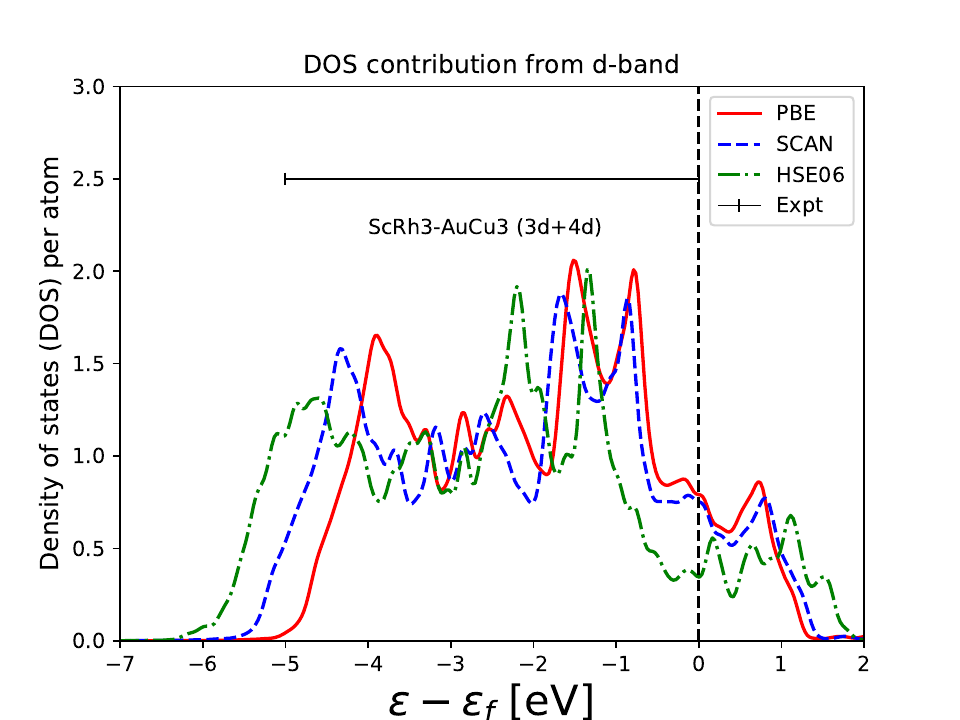}
   		\caption{ScRh$_3$ \cite{ScRh3}}
   		\label{fig:scrh3}
   	\end{subfigure}%
	\caption{The estimated valence d-band density of states of alloys with metals having partially-filled/partially-filled d-band configuration compared with valence d-band ranges extracted from experimental X-ray photoemission spectra or ultraviolet photoemission spectra (denoted by horizontal solid line). References are given in sub-captions. $\epsilon_f$ is the Fermi-level.}
	\label{fig:pf-pf2}
\end{figure}

In Figure~\ref{fig:pf-pf2}, we compare the PDOS of HfPt (Fig.~\ref{fig:hfpt}) and ScRh$_3$ (Fig.~\ref{fig:scrh3}) obtained using various methods. Unfortunately, we could not obtain experimental results for HfPt for comparison. But, the XPS valence d-band spectra of ScRh$_3$ is available \cite{ScRh3}. It has a d-band that ranges from the Fermi-level to around $\sim$ 5 eV below the Fermi-level \cite{ScRh3}, and it has a shape like that of YPd$_3$. As expected, HSE06 PDOS is blue-shifted away from the experimental range, while PBE and SCAN yield similar PDOS compared to the experiment. However, SCAN overestimates the experimental Rh's 4d band width by $\sim$ 1 eV, while the overestimation is only about $\sim$ 0.5 eV in the case of PBE. This result leads to accurate formation energy for the PBE, while SCAN overshoots the experimental value by $\sim$ 230 meV.\\

Though all DFT calculated d-band widths of hafnium and platinum metals are close to each other and agree with the experiment, we expect that only the SCAN calculated PDOS of HfPt should be close to that of an experimental result if available, as its formation energies are close to the experimental value.   

\section{Failure of RPA and beyond RPA correction}
Earlier, we have compared the PBE and SCAN results with HSE06, and established a connection between an accurate prediction of formation energies of alloys and their electronic properties. Here, we discuss the results obtained using the nonlocal random phase approximation (RPA). The RPA calculated formation energies of binary alloys in the B2 phase are tabulated in Table~\ref{table1} and compared with other semilocal and hybrid functionals as presented in Figure~\ref{fig2}. RPA provides accurate formation energies of CF-CF alloys with a lower energy ($\sim \hspace{0.2cm} <$ 130 meV) such as for AgZn, AgCd, and CuZn. This is consistent with the previous results for copper-gold alloys, which have formation energies less than 100 meV \cite{NABR19}. Other CF-CF alloys such as CuPd, AuCd, and AuZn have deviation ranges from 35$-$50 meV, which is significant as compared to experimental values. Furthermore, errors in RPA enormously increase when an alloy consists of partially-filled d-band metals. In general, it overestimates the experimental formation energies up to the maximum of $\sim$ 350 meV for PtSc.\\

Though RPA works reasonably well in predicting the cohesive energies of transition metal bulk comparable to PBE \cite{SGKMK13} with a mean absolute error $\sim$ 0.25 eV, there is a notable difference in the formation energies of alloys. In contrast to formation energy, the cohesive energy of transition metal alloys is a more difficult test for DFT functionals as it involves isolated transition metal atoms \cite{CT09}. Therefore, more error cancellation is expected for formation energy as both alloys and constituents have the same phase. Here, we have computed the RPA formation energies non-self-consistently using the ground-state PBE eigenstates and eigenvalues as reference. Therefore, we do not have RPA calculated partial density of states to compare with experiment. In RPA, the EXX energy ($E_{total} - E^{RPA}_c$) is one-electron self-interaction free, but the correlation energy $E^{RPA}_c$ suffers from self-correlation error due to the absence of the dynamical exchange-correlation kernel ($f_{xc}(q,\omega)$) \cite{EBF12,RRJ12}. RPA provides a good description of the long-range electron-electron correlation required to describe van der Waals interactions \cite{HK08,SHS10,BGKN12,NRB18}. However, the short-range electron-electron correlation is not properly described with bare RPA ($f_{xc} \rightarrow 0$) which yields too negative correlation energies (by $\sim$ 0.4 eV/electron) for the uniform electron gas in the metallic range densities \cite{LGP00,RNPP20}. Restoring the non-local exchange-correlation kernel can improve the repulsive short-range correlation, thereby giving an exact result \cite{LGP00, OT12, OT13, OT14,RNPP20}. We suspect that the too-negative formation energies (compared to experiment) by RPA for alloys with partially-filled d-band metals may be related to too-low correlation energy due to an imperfect description of the repulsive short-range electron-electron correlation. Consequently, we computed the formation energies of PtSc and HfOs using the renormalized adiabatic LDA (rALDA) \cite{OT12,OT13} and renormalized adiabatic PBE (rAPBE) \cite{OT14} kernels and tabulated them in Table~\ref{table3} (see Supplementary Tables S1 and S2 for input parameters of calculations).\\

\begin{table}
	\caption{Formation energies (eV) from kernel-corrected RPA for PtSc and HfOs. Improving the short-range correlation in RPA can improve the formation energies of intermetallics where RPA fails severely.}
	\begin{tabular}{|l|r|r|r|r|}
		\hline
		& \multicolumn{1}{l|}{RPA} & \multicolumn{1}{l|}{rALDA} & \multicolumn{1}{l|}{rAPBE} & \multicolumn{1}{l|}{Experiment} \\ \hline
		HfOs & -0.667 & -0.642 & -0.612 & -0.482 $\pm$ 0.052 \cite{IW18,MG98}\footnote{It is mentioned in the reference that HfOs alloy sample was not completely pure. It had small amounts of Hf$_{54}$Os$_{17}$ and relatively important quantities of unreacted Os. Therefore, the true result should be more exothermic than -0.518 eV \cite{MG98}.} \\ 
		PtSc & -1.440 & -1.317 & -1.257 & -1.086 $\pm$ 0.056 \cite{SK91,GK01} \\ \hline
	\end{tabular}
	\label{table3}
\end{table}

The rALDA kernel is obtained by using a (local) truncation of ALDA kernel for wavevector q $>$ 2$k_F$, where $k_F = (3 \pi^2 * density)^{1/3}$ is the Fermi wavevector \cite{OT12,OT13}. Also, rAPBE is similar to rALDA, except it also includes a PBE-like gradient correction \cite{OT14}. A large overestimation of formation energy from RPA can be reduced a little using rALDA. Further, rAPBE improves the RPA formation energies by a significant amount of $\sim$ 200 meV in the case of PtSc, closer to the experimental value. More corrections in the RPA formation energies can be expected when using a more exact uniform electron gas kernel, such as modified-CP07 (MCP07) \cite{RNPP20}.

\section{Conclusions}
We performed DFT calculations to compute the ground-state equilibrium properties of intermetallic alloys. Many of earlier studies \cite{ZKW14,NABR19} argued that the nonlocality is essential for accurate formation energies. However, those studies covered only a narrow range of weakly-bonded compounds having completely-filled d-band metals. Our assessment includes a broad range of binary alloys, which also include partially-filled d-band metals. We found that the nonlocality is not always useful for formation energies.  Instead, a PBE-like exchange-correlation can yield accurate results compared to the experiment, especially for strongly-bonded intermetallic alloys having partially-filled d-band metals.\\

Based on the observations, we classified intermetallic alloys into three categories based on their d-band filling combinations, e.g., completely-filled / completely-filled (CF-CF), completely-filled / partially-filled (CF-PF), and partially-filled / partially-filled (PF-PF). The formation energies usually increase in the order of CF-CF $<$ CF-PF $<$ PF-PF. As previously discussed, the nonlocal functionals HSE06 and RPA give accurate formation energies of CF-CF alloys, while PBE-GGA yields better results in the case of CF-PF and PF-PF alloys. Therefore, we suggest using a more PBE-like exchange-correlation for strongly-bonded alloys having a partially-filled d-band metal, while the nonlocality is necessary to capture the energy differences in the case of weakly-bonded alloys with completely-filled d-band transition metals. The difficulties to incorporate a delicate balance between the two such extreme cases make it arduous for any DFT approximation to provide the accurate equilibrium properties of a wide range of alloys. Nevertheless, a meta-GGA could be a natural trade-off, as it contains some nonlocality due to the kinetic energy density, while it still maintains its status as a semilocal approximation.\\

Besides, we also established a one-to-one correspondence between formation energies and electronic properties by estimating the d-band contribution to valence density of states (DOS). In other words, the functional which predicts accurate DOS of alloys and metals simultaneously also agrees with the experimental formation energies. The PBE-GGA underestimates the d-band range of completely-filled transition metals and their alloys while it provides electronic DOS often similar or better than HSE06 in the case of partially-filled d-band metals and their alloys. Contrarily, SCAN often improves on PBE calculated d-band centroid and hence the d-band range of many alloys and bulk metals, but it does not share a similar success as that of PBE. It may be due to a lack of required error cancellation in the SCAN meta-GGA that should be expected due to the same phase of the alloy and its constituents.\\

The state-of-the-art RPA, which can describe different bonding situations often much better than semilocal and hybrid functionals, severely fails for the formation energies of intermetallic alloys with partially-filled d-band metals. It significantly overestimates the formation energies. This may be related to the too-negative correlation energies within the metallic densities due to the incomplete description of repulsive short-range electron-electron correlations. Restoring the nonlocal exchange-correlation kernel rAPBE improves the formation energy of PtSc by $\sim$ 200 meV, which is substantial.
 
\section{Acknowledgements}
N.K.N. and A.R. acknowledge support by the National Science Foundation under Grant No. DMR-1553022. Computational support was provided by Temple University's HPC resources and thus was supported in part by the National Science Foundation through major research instrumentation grant number 1625061 and by the US Army Research Laboratory under contract number W911NF-16-2-0189. The work of S.A. was supported by the U.S. Department of Energy, Office of Sciences, Office of Basic Energy Sciences as part of the Computational Chemical Sciences Program under Award No. DE-SC0018331. The work of B.N. was supported as part of the Center of Complex Materials from First Principles (CCM), an Energy Frontier Research Center funded by the U.S. Department of Energy (DOE), Office of Science, Basic Energy Sciences (BES), under Award No. DE-SC0012575.     \\

\section{Supporting Information for ``Formation energy puzzle in intermetallic alloys: Random phase approximation fails to predict accurate formation energies"}
\begin{table}
	\renewcommand\thetable{S1}
	\caption{Converged parameters for elemental bulk calculations. The kernel-corrected or beyond RPA calculations were performed at RPA Ecut and a fixed K-mesh of 16x16x16. Also, the response function is computed using only one cutoff of 300 eV without extrapolation.}
	\begin{tabular}{|l|cc|cc|cc|}
		\hline
		& \multicolumn{1}{c}{HSE06} &  & \multicolumn{1}{l}{EXX} &  & \multicolumn{1}{l}{RPA} &  \\ \hline
		Elements & \multicolumn{1}{l}{Ecut (eV)} & K-mesh & \multicolumn{1}{l}{Ecut (eV)} & K-mesh & \multicolumn{1}{l}{Ecut (eV)} & K-mesh \\ \hline
		Sc & 400 & 13x13x13 & 700 & 16x16x16 & 600 & 18x18x18 \\ 
		Cu & 400 & 12x12x12 & 1000 & 16x16x16 & 800 & 16x16x16 \\ 
		Zn & 400 & 15x15x15 & 800 & 18x18x18 & 800 & 18x18x18 \\ 
		Y & 400 & 15x15x15 & 700 & 10x10x10 & 500 & 18x18x18 \\ 
		Rh & 400 & 13x13x13 & 700 & 16x16x16 & 600 & 18x18x18 \\ 
		Pd & 400 & 13x13x13 & 700 & 19x19x19 & 800 & 23x23x23 \\ 
		Ag & 350 & 12x12x12 & 600 & 16x16x16 & 600 & 16x16x16 \\ 
		Cd & 400 & 13x13x13 & 700 & 18x18x18 & 500 & 21x21x21 \\ 
		Hf & 400 & 13x13x13 & 700 & 16x16x16 & 600 & 19x19x19 \\ 
		Os & 400 & 15x15x15 & 700 & 19x19x19 & 600 & 20x20x20 \\ 
		Pt & 400 & 14x14x14 & 800 & 18x18x18 & 600 & 20x20x20 \\ 
		Au & 350 & 14x14x14 & 1000 & 15x15x15 & 800 & 16x16x16 \\ 
		Ti & 400 & 13x13x13 & \multicolumn{1}{l}{} &  & \multicolumn{1}{l}{} &  \\ \hline
	\end{tabular}
\end{table}

\begin{table}
	\renewcommand\thetable{S2}
	\caption{Converged parameters for alloys. The kernel-corrected or beyond RPA calculations were performed at RPA Ecut and a fixed K-mesh of 16x16x16. Also, the response function is computed using only one cutoff of 300 eV without extrapolation.}
	\begin{tabular}{|l|rl|rl|rl|}
		\hline
		& \multicolumn{1}{l}{HSE06} &  & \multicolumn{1}{l}{EXX} &  & \multicolumn{1}{l}{RPA} &  \\ \hline
		Alloys & \multicolumn{1}{l}{Ecut (eV)} & K-mesh & \multicolumn{1}{l}{Ecut (eV)} & K-mesh & \multicolumn{1}{l}{Ecut (eV)} & K-mesh \\ \hline
		AgZn & 450 & 14x14x14 & 800 & 17x17x17 & 700 & 20x20x20 \\ 
		AgCd & 450 & 14x14x14 & 800 & 16x16x16 & 800 & 18x18x18 \\ 
		CuZn & 450 & 14x14x14 & 1000 & 20x20x20 & 800 & 19x19x19 \\ 
		CuPd & 450 & 14x14x14 & 900 & 16x16x16 & 800 & 18x18x18 \\ 
		AuCd & 450 & 15x15x15 & 800 & 18x18x18 & 700 & 18x18x18 \\ 
		CuY & 450 & 14x14x14 & 900 & 18x18x18 & 700 & 18x18x18 \\ 
		AuZn & 450 & 15x15x15 & 900 & 16x16x16 & 700 & 18x18x18 \\ 
		ScAg & 450 & 14x14x14 & 700 & 18x18x18 & 700 & 18x18x18 \\ 
		AgY & 450 & 14x14x14 & 800 & 19x19x19 & 700 & 18x18x18 \\ 
		HfOs & 450 & 14x14x14 & 700 & 16x16x16 & 600 & 16x16x16 \\ 
		AuSc & 450 & 15x15x15 & 900 & 18x18x18 & 700 & 19x19x19 \\ 
		RhY & 450 & 15x15x15 & 700 & 16x16x16 & 600 & 18x18x18 \\ 
		ScPd & 450 & 16x16x16 & 700 & 21x21x21 & 500 & 18x18x18 \\ 
		ScRh & 450 & 14x14x14 & 700 & 16x16x16 & 500 & 18x18x18 \\ 
		PtSc & 450 & 17x17x17 & 800 & 17x17x17 & 800 & 15x15x15 \\ 
		HfPt & 450 & 14x14x14 & 700 & 18x18x18 & 600 & 18x18x18 \\ 
		AuTi & {400} & 11x11x11  & \multicolumn{1}{l}{} &  & \multicolumn{1}{l}{} &  \\ 
		ScRh3 & 400 & 12x12x12 & \multicolumn{1}{c}{} &  & \multicolumn{1}{c}{} &  \\ 
		YPd3 & 400 & 14x14x14 & \multicolumn{1}{c}{} &  & \multicolumn{1}{c}{} &  \\ 
		ScPt3 & 400 & 12x12x12 & \multicolumn{1}{c}{} &  & \multicolumn{1}{c}{} & \\ \hline
	\end{tabular}
\end{table}

\subsection{Equilibrium volume} 
In this section, we present the comparison between estimated and experimental equilibrium volumes for both elemental bulk constituents and alloys in Tables S3 and S4 respectively. As expected, PBE overestimated the equilibrium volume in most cases with excellent results for 3d and partially-filled 5d transition metals within $-$0.2 to 3.0 \%, while the results deviate more with the experiments for 4d and completely-filled 5d elements up to 6 \%. Contrary to PBE, SCAN provides accurate results for completely-filled 4d and 5d elements with an error of less than one percent. HSE06 hybrid provides more improvement keeping the trend similar to PBE, while the RPA further improves on it. For alloys, SCAN provides notable correction to PBE results, except for CuZn and HfPt where one can observe that the error is propagated from the discrepancies of their elemental bulk results: Cu ($-$3.8 \%), Zn ($-$7.8 \%), Hf ($-$4.0 \%), and Pt ($-$4.0 \%). Once again, RPA delivers accurate results for these alloys similar to their bulk constituents mostly within two percent error compared to experiment.\\ 

We calculated the mean absolute error (MAE) and mean absolute percentage error (MAPE) for a better understanding of performances of various functionals predicting equilibrium volumes of the system considered in the present study. As expected, we can observe that the error decreases from semilocal DFT functionals to nonlocal functionals in both types of systems. The MAE for elemental bulk systems is in the order of RPA $<$ HSE06 $<$ PBE $<$ SCAN, while MAPE shows a trend of RPA $<$ HSE06 $<$ SCAN $<$ PBE. On the other hand, the SCAN functional performs better than HSE06 for alloy systems with MAPE in the order of RPA $<$ SCAN $<$ HSE06 $<$ PBE. 
\begin{table*}
	\renewcommand\thetable{S3}
	\caption{Equilibrium volumes for constituent elemental bulk transition metals (\AA$^3$). }
	\resizebox{1.\columnwidth}{!}{%
		\begin{tabular}{cccccccccc}
			\hline
			Element & \multicolumn{1}{l}{PBE} &\% Error &\multicolumn{1}{l}{SCAN} & \% Error &HSE06& \% Error & \multicolumn{1}{l}{RPA}& \% Error& \multicolumn{1}{l}{Experiment \cite{OKS96}} \\ \hline
			Sc & 49.420 &0.1 & 49.944 &1.1 &50.655  &2.6  &49.390 &0.02  &49.38 \\ 
			Cu & 12.004 &1.6  & 11.361 &-3.8 &12.021 &1.8 &\multicolumn{1}{l}{11.958} &1.3 & {11.81} \\ 
			Zn & 30.392 &-0.2 & 28.072 &-7.8 & 30.406 &-0.1 & 29.620 &-2.7 & {30.44} \\ 
			Y & 65.489 &-0.8 & 67.409 &2.1 & 67.200 &1.8 & \multicolumn{1}{l}{64.872} &-1.7 & 66.02 \\ 
			Rh & 14.050 &2.2 & 13.515 &-1.7 & 13.54 &-1.5 & 13.930 &1.3 & 13.75 \\ 
			Pd & 15.341 &4.2 & 14.766 &0.2 & 15.018 &2.0 & 14.650 &-0.5 & 14.73 \\ 
			Ag & 17.827 &4.5 & 17.043 &-0.1 & 17.798 &4.3 & 17.090 &0.2 & 17.06 \\ 
			Cd & 45.886 &6.3 & 43.075 &-0.2 & 44.631 &3.4 & 42.410 &-1.8 & {43.17} \\ 
			Hf & 44.896 &0.3 & 42.971 &-4.0 & 44.671 &-0.2 & 45.440 &1.5 & 44.76 \\ 
			Os & 28.538 &2.0 & 27.341 &-2.3 & 27.661 &-1.1 & 28.060 &0.3 & 27.98 \\ 
			Pt & 15.610 &3.0 & 14.537 &-4.0 & 15.248 &0.7 & 15.200 &0.3 & 15.15 \\ 
			Au & 17.961 &5.8 & 17.103 &0.7 & 17.595 &3.7 & \multicolumn{1}{l}{17.868} &5.2 & {16.98} \\
			MAE &  & 0.62 &  & 0.69  &  & 0.53  & & 0.4 & \\
			MAPE (\%) & & 2.6  & & 2.3 & & 1.8 & & 1.4  & \\
			\hline
		\end{tabular}
	}
	\label{tab:vol1}
\end{table*}

\begin{table*}
	\renewcommand\thetable{S4}
	\caption{Equilibrium volumes for alloys (\AA$^3$); Alloys are presented with an increasing experimental formation energies (see Table V); \% Error means percentage error.}
	\resizebox{1.\columnwidth}{!}{%
		\begin{tabular}{lccccccccc}
			\hline
			Alloys & \multicolumn{1}{l}{PBE} & \% Error & \multicolumn{1}{l}{SCAN} & \% Error & HSE06 & \% Error & \multicolumn{1}{l}{RPA} &\% Error & \multicolumn{1}{l}{Experiment \cite{OKS96}} \\ \hline
			AgZn & 32.346 & 2.6 & 30.428 & -3.4 & 32.019 & 1.6 & 31.242 &-0.9  & 31.519 \\ 
			AgCd & 39.005 & 4.0  & 37.053 &-1.2  & 38.405  &2.4 & 36.886 &-1.6 & 37.494 \\ 
			CuZn & 26.172 & 1.1 & 24.493 &-5.4 & 26.119 &0.9 & 25.707 &-0.7 & 25.882 \\ 
			CuPd & 27.449 & 3.6 & 26.257 &-0.9 & 27.153 &2.5 & 26.946 &1.7 & 26.490 \\
			AuCd & 39.059 & 6.3 & 37.109 &1.0 & 38.176 &3.9 & 37.591 &2.4 & 36.727 \\
			CuY & 42.264 &0.5 & 41.254 &-1.9 & 42.692 &1.6 & 41.663 &-0.9 & 42.035 \\ 
			AuZn & 32.397 & 4.1 & 30.432 &-2.2 & 31.856 &2.4 & 31.847 &2.4 & 31.107 \\ 
			ScAg & 40.703 & 2.5 & 39.335 &-1.0 & 40.537 &2.1 & 39.574 &-0.4 & 39.722 \\ 
			AgY & 48.381 & 2.8& 46.930 &-0.3 & 48.398 &2.9 & 47.002 &-0.1 & 47.046 \\ 
			HfOs & 34.504 &1.5 & 33.229 &-2.2 & 33.795 &-0.5 & 34.570 &1.7 & 33.981 \\ 
			AuSc & 39.525 & 3.3& 38.014 &-0.7 & 39.118 &2.2 & 38.943 &1.8 & 38.273 \\ 
			RhY & 40.449 &2.3 & 39.432 &-0.3 & 39.591 &0.1 & 40.254 &1.8 & 39.547 \\ 
			ScPd & 36.501 & 3.4 & 35.450 &0.5 & 36.073 &2.2 & 35.989 &2.0 & 35.288 \\ 
			ScRh & 33.775 &2.7 & 32.850 &-0.1 & 33.096 &0.6 & 33.703 &2.5 & 32.891 \\ 
			PtSc & 36.109 &3.5 & 34.722 &-0.5 & 35.634 &2.1 & 35.866 &2.8 & 34.902 \\ 
			HfPt & 36.846 &-3.1 & 35.130 &-7.6 & 36.244 &-4.6 & 36.637 &-3.6 & 38.011 \\

			MAE &  &1.06 & & 0.62 & & 0.74 & & 0.59  & \\
			MAPE (\%) & &2.97& & 1.81 & & 2.05 & & 1.70  & \\
			\hline
		\end{tabular}
	}
	\label{tab:vol2}
\end{table*}

\subsection{Bulk Moduli}
Bulk moduli are more difficult to predict than equilibrium volumes, as they require an accurate prediction of the equation of state. We computed bulk moduli by fitting the Birch-Murnaghan equation-of-state to the energy-volume data and compared them with available experimental results in Tables S5 and S6. The PBE-GGA provides reasonably accurate bulk moduli of 3d metals, while the errors are enormous for 4d and 5d elements. On the other hand, SCAN meta-GGA significantly improves upon PBE, and it is more accurate in most cases except copper, zinc, and platinum. The error for SCAN is a maximum at about 42 \% in zinc, and it is consistent with the error in its equilibrium volume. Both HSE06 and RPA have a mixed performance on bulk moduli with lower overall mean absolute error (MAE) and mean absolute percentage error (MAPE) than semilocal results. We also compared computed bulk moduli of intermetallic alloys with available experimental results. Unlike constituent elemental bulk metals, all DFT approximations studied here predict reasonable bulk moduli. For example, the estimated bulk modulus of CuZn alloy agrees with the experimental value, though, the bulk moduli of zinc significantly deviates from the experiment.

\begin{table*}
	\renewcommand\thetable{S5}
	\caption{Bulk moduli for the constituent elemental bulk transition metals (GPa); Experimental bulk moduli are taken from Ref~\cite{JLSVL14}.}
	\resizebox{1.\columnwidth}{!}{%
		\begin{tabular}{cccccccccc}
			\hline
			Element & \multicolumn{1}{l}{PBE} &\% Error & \multicolumn{1}{l}{SCAN} & \% Error& HSE06 & \% Error & \multicolumn{1}{l}{RPA} & \% Error & \multicolumn{1}{l}{Experiment \cite{JLSVL14}} \\ \hline
			Sc & 53.6 &-3.5 & 55.7 &0.2 & 56.1 &0.9 & 64.0 &15.1 & 55.6 \\ 
			Cu & 137.9 &-1.7 & 156.2 &11.3 & 125.8 &-10.3 & \multicolumn{1}{l}{144.7} &3.1 & {140.3} \\ 
			Zn & 73.4 &5.3 & 99.4 &42.6 & 74.8 &7.3 & 84.2 &20.8 & 69.7 \\ 
			Y & 39.9 &-4.3 & 41.2 &-1.3 & 40.3 &-3.4 & \multicolumn{1}{l}{45.7} &9.6 & 41.7 \\ 
			Rh & 254.8 &-11.8 & 289.5 &0.3 & 288.2 &-0.2 & 311.4 &7.9 & 288.7 \\ 
			Pd & 164.4 &-15.9 &  206.5 &5.7 &  171.9 &-12.0 & 231.6 &18.5 & 195.4 \\ 
			Ag & 91.0 &-12.3 & 110.5 &6.5 & 85.8 &-17.3 & 115.3 &11.1 & 103.8\\ 
			Cd & 43.2 &-19.7 & 59.3 &10.2 & 52.4 &-2.6 & 62.9 &16.9 & 53.8 \\ 
			Hf & 108.3 &-1.3 & 115.1 &4.9 & 114.2 &4.1 & 114.1 &4.0 & 109.7 \\ 
			Os & 403.8 &-4.9 & 459.6  &8.2 &460.7  &8.5 & 433.2 &2.0 & 424.6 \\ 
			Pt & 247.3 &-13.0 & 331.5 &16.6 & 267.4 &-5.9 & 298.8 &5.1 & 284.2 \\ 
			Au & 138.9 &-20.5 & 166.5 &-4.8 & 144.6 &-17.3 & \multicolumn{1}{l}{176.7} &1.1 & {174.8}  \\ 
			MAE (GPa) &  &16.1 &  &13.9 &  &12.7 &  &11.7 & \\
			MAPE (\%) & &9.5 &  &9.4 &  &7.5 &  &9.6 & \\\hline
		\end{tabular}
	}
	\label{fig3}
\end{table*}

\begin{table*}
	\renewcommand\thetable{S6}
	\caption{Bulk moduli for the alloys (GPa); The alloys are presented in order of increasing experimental formation energies (see Table V).}
	\begin{tabular}{lrrrrr}
		\hline
		BM & \multicolumn{1}{l}{PBE} & \multicolumn{1}{l}{SCAN} & HSE06 & \multicolumn{1}{l}{RPA} & Experiment \\ \hline
		AgZn & 90.5 & 112.3 & 92.3 & 107.8 &  \\ 
		AgCd & 73.8 & 90.6 & 76.9 & 95.2 &  \\ 
		CuZn & 112.8 & 141.3 & 110.1  & 118.4 & \multicolumn{1}{r}{116 \cite{L49}} \\ 
		CuPd & 151.6 & 176.8 & 147.8  & 169.3 &  \\ 
		AuCd & 90.5 & 110.5 & 98.8 & 116.6 & 100 \cite{NMMMK73,GE09} \\ 
		CuY & 69.3 & 74.6 & 67.0 & 77.1 & \multicolumn{1}{r}{70.1 \cite{MYLHGR04}} \\ 
		AuZn & 115.6 & 143.6 & 122.6 & 137 & \multicolumn{1}{r}{131} \cite{Z56} \\ 
		ScAg & 79.7 & 94.4 & 83.4 & 99.3 &  \\ 
		AgY & 66.1 & 75.1 & 66.6 & 79.7 & \multicolumn{1}{r}{70.1 \cite{MYLHGR04}} \\ 
		HfOs & 224.9 & 250.0 & 245.5 & 235.4 &  \\ 
		AuSc & 103.3 & 123.1 & 112.5 & 125.2 &  \\
		RhY & 108.8 & 121.4 & 117.1 & 121.8 &  \\ 
		ScPd & 110.7 & 124.8 & 116.5 & 127.6 &  \\ 
		ScRh & 138.3 & 154.9 & 152.0 & 154.6 &  \\ 
		PtSc & 136.4 & 157.7 & 147.5 & 155.1 &  \\ 
		HfPt & 179.2 & 209.1 & 192.0 & 191.5 &  \\ \hline

	\end{tabular}
	\label{fig4}
\end{table*}

\begin{figure}
	\renewcommand\thefigure{S1}
	\includegraphics[scale=0.5]{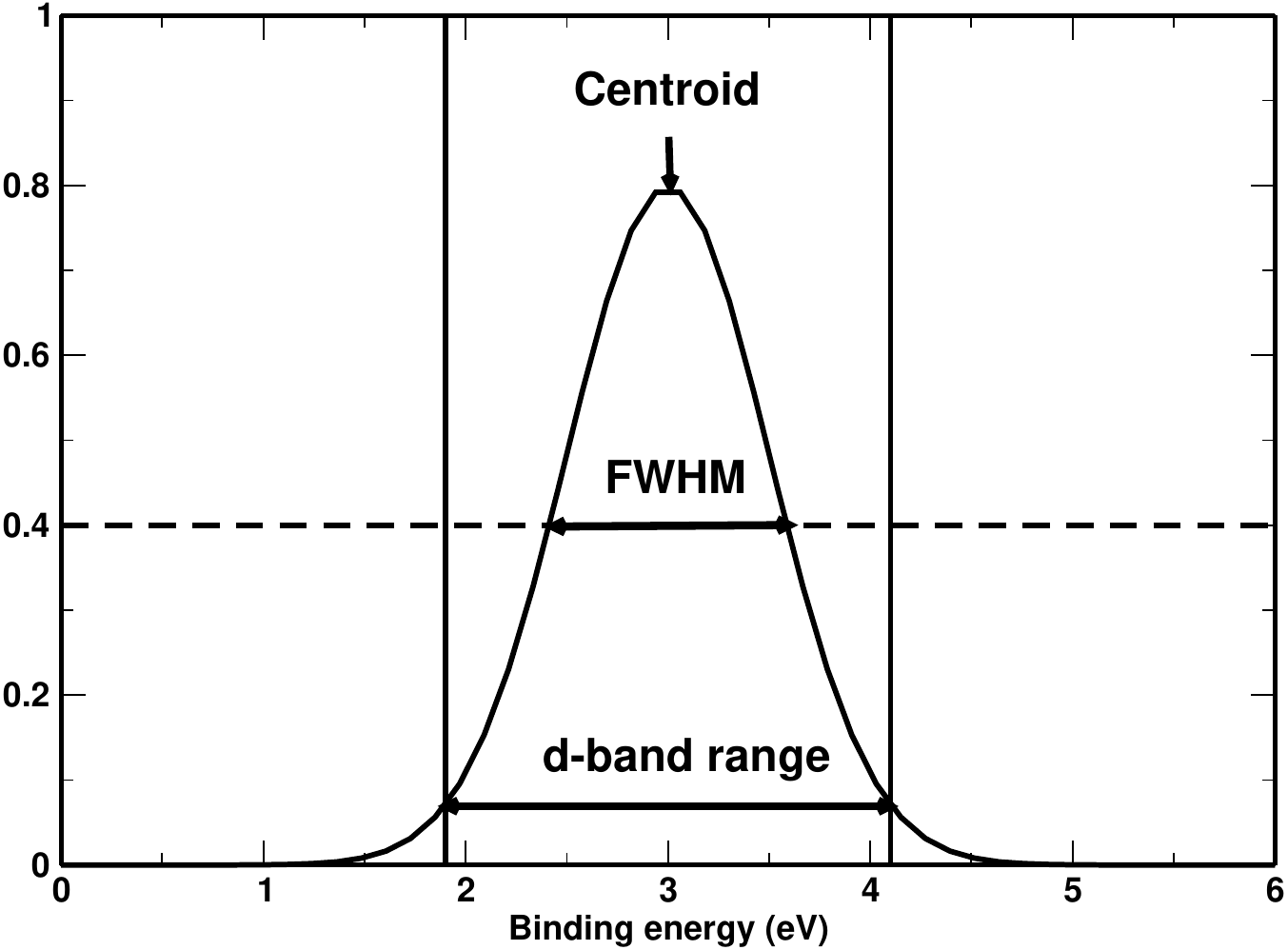}
	\caption{Schematic diagram for estimating the d-band range and FWHM from X-ray or ultraviolet photoemission spectra.}
\end{figure}

\pagebreak
\section{Experimental data for valence d-band:}
Both X-ray and ultraviolet photoemission spectra work via the photoelectric effect. When an atom absorbs a photon of energy $h\nu$, an electron in a core or valence state with binding energy $E_b$ is ejected with kinetic energy $E_k$ as
\begin{equation}
E_k = h\nu - E_b - \phi,
\end{equation}
\noindent where, $\phi$ is the work function and $E_b$ ($E - E_{Fermi}$) is the energy of the photoelectron with respect to the Fermi energy of the system. In principle, the photoemission spectrum can give DOS. However, one should be cautious with an exact comparison of theoretical DOS with experimental photoemission spectra (for exact features including satellite excitations and tail at high binding energy). The factors that play a crucial role in the difference between theoretical and experimental DOS could be the influence of matrix elements that govern the excitation of the photoelectrons and the photon energy itself used for the excitation process \cite{ScPdScAg}. Other factors may be the instrumental resolution and electron-electron scattering processes. Here, we are not interested in the exact features of the photoemission spectra.  Instead, the information about the valence d-band centroid (or binding energy) and range (or width) of the system is sufficient, as the previously-mentioned factors have an insignificant effect on it \cite{PdPt,Pd,ScRh3,Y}. Note that the term ``d-band range" is different than the full width at half maximum (FWHM) used in several references. The d-band range is measured after removing the extra asymptotic tail, while FWHM is measured at half of the maximum intensity as shown in Supplementary Figure S1. We have extracted the d-band centroid and the d-band range of both alloys and its constituents from experimental photoemission spectra, and tabulated them in Supplementary Table S7.

\begin{table}[htbp]
	\renewcommand\thetable{S7}
	\caption{The d-band centroid and d-band ranges with respect to Fermi-level (or below Fermi-level) extracted from experimental photoemission spectra; unit eV.}
	\begin{tabular}{cccc}
		\hline
		\hline
		Systems & d-band & \multicolumn{1}{c}{d-band centroid} & d-band range \\ \hline
		Cu & 3d & \multicolumn{1}{c}{3.0$-$3.5 } & 2.0$-$5.0 \cite{AuCuAg}, 1.5$-$4.5 \cite{CuZn} \\ 
		Zn & 3d & 10 & 8.5$-$11.5 \cite{CuZn,AuZn,AgZn} \\ 
		Pd & 4d & 1 & 0$-$5.5 \cite{Pd,PdPt} \\ 
		Ag & 4d & \multicolumn{1}{c}{} & 3.9$-$7.4 \cite{AuCuAg,AgZn} \\ 
		Cd & 4d & 11 & 9.0$-$13.0 \cite{AgCd,AgCd2} \\ 
		Au & 5d & \multicolumn{1}{c}{} & 2.0$-$8.0 \cite{AuCuAg,AuZn} \\ 
		Sc & 3d & 0.2 & 0.0$-$1.5 \cite{ScPdScAg} \\ 
		Y & 4d & \multicolumn{1}{c}{} & 0$-$2.0 \cite{Y} \\ 
		Rh & 4d & \multicolumn{1}{c}{1.3$-$1.5} & 0$-$5.0\cite{RhAg,Rh} \\ 
		Hf & 5d & 0.9 & 0$-$4.0 \cite{Hf} \\ 
		Os & 5d & 3 & 0$-$8.0 \cite{Rh} \\ 
		Pt & 5d & 1.6 & 0$-$8.0 \cite{PdPt} \\ 
		\hline
		AgZn & 4d-3d &-- &4.0$-$11.0 \cite{AgZn} \\
		AgCd & 4d-4d &-- &4.5$-$11.5 \cite{AgCd,AgCd2} \\
		CuZn &3d-3d &-- &2.9$-$10.7 \cite{CuZn} \\
		AuZn &5d-3d &-- &3.7$-$11.5 \cite{AuZn} \\
		AgSc &4d-3d &-- &0$-$7.0 \cite{ScPdScAg} \\
		ScPd &3d-4d &-- &0$-$5.0 \cite{ScPdScAg} \\
		YPd$_3$& 4d-4d &-- &0$-$4.8 \cite{YPd3,Pd}  \\
		ScRh$_3$ &3d-4d &-- &0$-$5.0 \cite{ScRh3} \\
		
		\hline
		\hline
	\end{tabular}
	\label{}
\end{table}



\end{document}